\documentclass[11pt]{scrartcl}

\usepackage[utf8]{inputenc} 
\usepackage{mathtools, amsfonts, amssymb, breqn}
\usepackage{amsthm}

\usepackage[british]{babel}
\usepackage{graphicx, hyperref, tabularx, abstract, makecell,ragged2e}

\usepackage[format=plain]{caption}
\DeclareCaptionFormat{cont}{#1 (cont.)#2#3\par}

\usepackage[outdir=./]{epstopdf}   

\usepackage{cellspace, hhline} 

\usepackage{subfig}
\usepackage{xcolor-solarized}
\usepackage{todonotes}
\usepackage{overpic}

\usepackage{enumitem}
\setdescription{itemsep=5pt,parsep=0pt,leftmargin=1.55cm,font=\normalfont}

\usepackage[]{geometry}
\newcommand{\R}{\ensuremath{\mathbb{R}}}
\newcommand{\C}{\ensuremath{\mathbb{C}}}

\newcommand{\dif}{{\operatorname{d}}}

\usepackage{cellspace, hhline}

\numberwithin{equation}{section}
\numberwithin{figure}{section}
\numberwithin{table}{section}

\begin{document}
	\author{L. Eigentler and J. A. Sherratt}
	
	\newcommand{\Addresses}{{
			\bigskip
			\footnotesize
			
			L. Eigentler (corresponding author),  Maxwell Institute for Mathematical Sciences, Department of Mathematics, Heriot-Watt University, Edinburgh EH14 4AS, United Kingdom\par\nopagebreak
			\textit{E-mail address}: \texttt{le8@hw.ac.uk}
			
			\medskip
			
			J. A. Sherratt,  Maxwell Institute for Mathematical Sciences, Department of Mathematics, Heriot-Watt University, Edinburgh EH14 4AS, United Kingdom\par\nopagebreak
			\textit{E-mail address}: \texttt{J.A.Sherratt@hw.ac.uk}

	}}

\title{Analysis of a Model for Banded Vegetation Patterns in Semi-Arid Environments with Nonlocal Dispersal}

\date{}

\maketitle

\begin{abstract}
	Vegetation patterns are a characteristic feature of semi-arid regions. On hillsides these patterns occur as stripes running parallel to the contours. The Klausmeier model, a coupled reaction-advection-diffusion system, is a deliberately simple model describing the phenomenon. In this paper, we replace the diffusion term describing plant dispersal by a more realistic nonlocal convolution integral to account for the possibility of long-range dispersal of seeds. Our analysis focuses on the rainfall level at which there is a transition between uniform vegetation and pattern formation. We obtain results, valid to leading order in the large parameter comparing the rate of water flow downhill to the rate of plant dispersal, for a negative exponential dispersal kernel. Our results indicate that both a wider dispersal of seeds and an increase in dispersal rate inhibit the formation of patterns. Assuming an evolutionary trade-off between these two quantities, mathematically motivated by the limiting behaviour of the convolution term, allows us to make comparisons to existing results for the original reaction-advection-diffusion system. These comparisons show that the nonlocal model always predicts a larger parameter region supporting pattern formation. We then numerically extend the results to other dispersal kernels, showing that the tendency to form patterns depends on the type of decay of the kernel.

\end{abstract}


\section{Introduction}\label{sec:Intro}


\subsection{Ecological Background}\label{sec:Intro Ecological Background}
Semi-arid environments are regions in which the level of rainfall is below a certain threshold, dependent on the mean temperature and spread of rainfall across the year \cite{Koeppen1936, Peel2007}, creating a hostile environment for vegetation as plants compete for water. A characteristic feature of many of these semi-arid environments is self-organised patterns of vegetation. These occur due to a scale-dependent feedback, which is caused by the modification of the soil by the existing plants, creating a more favourable environment on a short range and the competition for water on a longer spatial distance \cite{Rietkerk2008}. On gentle slopes of a few percent gradient (0.2\% to 2\% \cite{Valentin1999}) striped patterns occur along the contours of the hill. Being wide and with large distances between them, these stripes are extremely difficult to detect from the ground. They were therefore first discovered using aerial photography in the 1950s in British Somaliland (today Somalia) \cite{Macfadyen1950, Hemming1965}. Since then, striped patterns have been observed on slopes in the Chihuahuan Desert in Mexico and the US \cite{Cornet1988, Montana1990, Montana1992}, New South Wales in Australia \cite{Tongway1990, Dunkerley2002}, Niger and other countries in the African Sahel \cite{Thiery1995, Worrall1959, White1971} and many other regions as reviewed by \cite[Table 1 and Figure 3]{Valentin1999}. Many ecologists studying these patterns reported that the vegetation bands slowly move uphill \cite{Valentin1999, Worrall1959, Montana1992} with a migration speed varying between 0.2m and 1.5m per year \cite{Valentin1999}. They argue that the reason for this is that the rainwater, which often falls in form of torrential rain at irregular intervals \cite{Bromley1997}, runs off the bare ground to the uphill edge of the vegetation band below, where it can infiltrate the ground more easily, providing a more favourable environment for plant growth on the uphill edge than on the downhill edge \cite{White1971, Montana2001}. Other authors observed stationary patterns \cite{Dunkerley2002}, which they attribute to changes in the soil on bare ground that inhibits plant growth \cite{Dunkerley2002} and a skewed distribution of plant dispersal caused by seeds travelling downhill in the flow of the water \cite{Saco2007, Thompson2009}. A more recent survey confirms the occurrence of both upward migration and static vegetation bands, by comparing satellite data from spy satellites used during the Cold War to more recent data \cite{Deblauwe2010}. Studying these patterns is of crucial importance as changes in the width of and distance between vegetation stripes may be an indicator for an imminent and irreversible switch to desertification \cite{Kefi2007, Rietkerk2004}. 

The long timescale in the evolution of patterned vegetation and the inability to generate it in laboratory settings limit the availability of observed data. Instead various different theoretical models have been developed \cite{Borgogno2009}. These can be classified into two main groups; models based on plant to plant interactions, among other things including individual plant's morphology such as its root network and shading \cite{Gilad2004, Gilad2007, Lefever2009, Hardenberg2010} and models focusing on water redistribution. The latter class of models are based on the Klausmeier model \cite{Klausmeier1999}, on which we will focus here.


\subsection{The Models}\label{sec:Intro Models}

The nondimensionalised form of the Klausmeier model (see \cite{Klausmeier1999, Sherratt2005} for details on the nondimensionalisation) is the reaction-advection-diffusion system

\begin{subequations}
	\label{eq:Intro Klausmeier local}
	\begin{align}
	\frac{\partial u }{\partial t} &= \overbrace{u^2w}^{\text{plant growth}} - \overbrace{Bu}^{\text{plant loss}} + \overbrace{\frac{\partial^2 u}{\partial x^2}}^{\text{plant dispersal}}, \\
	\frac{\partial w }{\partial t} &= \underbrace{A}_{\text{rainfall}} - \underbrace{w}_{\text{evaporation}} - \underbrace{u^2w}_{\substack{\text{water uptake} \\ \text{by plants}}} + \underbrace{\nu\frac{\partial w}{\partial x}}_{\substack{\text{water flow}\\ \text{downhill}}} +  \underbrace{d\frac{\partial^2w}{\partial x^2}}_{\substack{\text{diffusion} \\ \text{of water}}}.
	\end{align}
\end{subequations}
Originally, this model did not include diffusion of water, but this term was added later and is now well established \cite{Stelt2013, Zelnik2013, Kealy2012, Siteur2014}. This extended Klausmeier model will be referred to as the ``local Klausmeier model'' throughout the text. In the model, $u(x,t)$ represents the plant density, $w(x,t)$ the water density, $t>0$ the time and $x \in \R$ the space, where the positive direction is in the uphill direction of a one-dimensional domain of constant gradient. The system assumes constant rainfall, proportionality of water density to evaporation \cite{Rodriguez-Iturbe1999, Salvucci2001} and correlation of plant growth to water uptake. The latter is assumed to be proportional to the water density and the plant density squared, because the water infiltration capacity of the soil depends on the presence of plants \cite{Valentin1999, Rietkerk2000}. The ground where vegetation stripes are situated is estimated to receive around 1.5 to 2.5 times as much water as the annual precipitation due to water running off the bare ground towards the vegetation stripes \cite{Cornet1988}. The parameters $A>0$, $B>0$, $\nu>0$ and $d>0$ represent rainfall, plant loss, the rate of the water flow in the downhill direction and the rate of water diffusion, respectively. Due to the nondimensionalisation they are however a combination of different ecological quantities. Parameter estimates are $A\in [0.1, 3]$, $B \in [0.05, 2]$ \cite{Klausmeier1999, Rietkerk2002} and $\nu = 182.5$ \cite{Klausmeier1999}. The large size of $\nu$ compared to the other parameters reflects the slow speed of plant dispersal compared to water flow, and it allows an analysis of patterned solutions of \eqref{eq:Intro Klausmeier local} by obtaining results for the model to leading order in $\nu$, such as by \cite{Sherratt2005, Sherratt2010, Sherratt2011, Sherratt2013III, Sherratt2013IV, Sherratt2013V}. The model is deliberately kept simple. There are however a wide range of systems all based on the Klausmeier model \eqref{eq:Intro Klausmeier local} that take into account variable precipitation \cite{Kletter2009} and grazing \cite{HilleRisLambers2001, Koppel2002} and models that distinguish between the surface water density and the water density in the soil \cite{HilleRisLambers2001, Rietkerk2002, Gilad2007}. 

In \eqref{eq:Intro Klausmeier local}, plant dispersal is modelled by a diffusion term. In reality, nonlocal processes are often involved, such as seed dispersal by wind or separated stages for plant growth and seed dispersal \cite{Pueyo2008}. This can be modelled by integrodifferential equations \cite{Allen1996, Powell2004}. To do this, the change of the plant density $u(x,t)$ at a point $x$ that was caused by diffusion is replaced by the convolution integral
\begin{align*}
\int_{-\infty}^{\infty} \phi(x-y)\left(u(y,t)-u(x,t)\right)\dif y.
\end{align*}
The kernel function $\phi(x,y)$ is a probability density function, describing the probability per unit length of seeds originating at the point $y$ being dispersed to point $x$ \cite{Pueyo2008}. This approach is not only used in modelling plant dispersal but can, among others, be considered to model dispersal in general competition models \cite{Hutson2003, Cosner2012} showing an evolutionary advantage of nonlocal dispersal under certain boundary conditions \cite{Kao2010}, or models describing a single species subject to a unidirectional flow \cite{Lutscher2005}. It is to assumed that seed dispersal only depends on the distance $x-y$ (i.e. assuming homogeneous and isotropic dispersion of seeds \cite{Mistro2005}). This kind of nonlocal seed dispersal is considered for example by \cite{Pueyo2008, Pueyo2010, Baudena2013} for modified versions of the Klausmeier model that consider soil water separately from surface water \cite{HilleRisLambers2001, Rietkerk2002}. Motivated by this, we will consider the ``nonlocal Klausmeier model'' 
\begin{subequations}
	\label{eq:Intro Klausmeier nonlocal}
	\begin{align}
	\frac{\partial u}{\partial t} &= u^2w - Bu + C\left(\int_{-\infty}^\infty \phi(x-y)u(y,t) \dif y - u(x,t)\right), \label{eq:Intro Klausmeier nonlocal plants}   \\
	\frac{\partial w}{\partial t} &= A-w-u^2w + \nu\frac{\partial w}{\partial x}+ d\frac{\partial^2w}{\partial x^2}.
	\end{align}
\end{subequations}
The dispersal coefficient $C>0$, which scales the convolution term, describes the plant's dispersal rate by taking into account the plant's fecundity, seed mortality and germination rate and seed establishment ability \cite{Pueyo2008}.  

If the kernel function $\phi(x)$ is decaying exponentially as $x \rightarrow \infty$, the local model can be obtained from the nonlocal model by setting $C = 2/\sigma(a)^2$, where $\sigma(a)$ denotes the standard deviation of the dispersal kernel with scaling parameter $a$, and taking the limit as $a\rightarrow \infty$. To show this, write $\phi(x) = a\varphi(ax)$. Then, the integral in the dispersal term can be transformed to
\begin{align*}
\int_{-\infty}^\infty \phi(x-y) u(y,t) \dif y = \int_{-\infty}^\infty \varphi(z) u\left(x-\frac{z}{a},t\right) \dif z,
\end{align*}
by using the change of variables $y = x-z/a$. Considering the Taylor expansion of $u(x-z/a,t)$ in $z/a$, an application of Watson's lemma (i.e. integrating term-wise) gives
\begin{multline}\label{eq: intro: convolution expansion}
\int_{-\infty}^\infty \phi(x-y) u(y,t) \dif y \\ = u(x,t) - \frac{1}{a} \frac{\partial u}{\partial x}(x,t) \int_{-\infty}^\infty \varphi(z) z \dif z + \frac{1}{2a^2}\frac{\partial^2 u}{\partial x^2}(x,t) \int_{-\infty}^\infty \varphi(z) z^2 \dif z +O\left(\frac{1}{a^3}\right).
\end{multline}
In this paper we will assume that the kernel $\phi$ is even with its mean located at $x=0$. Therefore the coefficient of the first order derivative in \eqref{eq: intro: convolution expansion} is zero and thus 
\begin{align*}
\int_{-\infty}^\infty  \phi(x-y) u(y,t) \dif y = u(x,t)  + \frac{\sigma(a)^2}{2}\frac{\partial^2 u}{\partial x^2}(x,t) +O\left(\frac{1}{a^3}\right),
\end{align*}
using $\varphi(x) = \phi(x/a)/a$ and the definition of the second moment of a probability distribution. Therefore, setting $C= 2/\sigma(a)^2$ gives
\begin{align*}
C\left( \int_{-\infty}^\infty \phi(x-y) u(y,t) \dif y - u(x,t)\right) = \frac{\partial^2 u}{\partial x^2}(x,t)+O\left(\frac{1}{a}\right) \rightarrow \frac{\partial^2 u}{\partial x^2}(x,t),
\end{align*} 
as $a \rightarrow \infty$. This limiting behaviour will allow us to make comparisons between the local and the nonlocal model. Two kernel functions for which the derivation above holds true are the Laplacian
\begin{align}\label{eq:Intro Laplacian kernel}
\phi(x) = \frac{a}{2}e^{-a|x|}, 
\end{align}
and the Gaussian distribution
\begin{align}\label{eq:Intro Gaussian kernel}
\phi(x) = \frac{a_g}{\sqrt{\pi}}e^{-a_g^2x^2},
\end{align}
where $x\in\R$, and $a, a_g >0$ are the scale parameters of the distributions, respectively. The Laplacian kernel corresponds to plants (seeds) dispersing as a random walk with individual plants (seeds) settling at different random times \cite{Neubert1995, Bullock2017}. One main goal of this paper is to investigate how a change in the width of the kernel affects the tendency to form patterns. Closely related to this, a second main aspect we will address in this paper is a comparison between different dispersal kernels. In particular we will show that the type of decay (i.e. exponential or algebraic) has an influence on the tendency to form patterns. The Laplacian kernel is not only biologically relevant \cite{Neubert1995, Bullock2017, Johnson1988} but also allows us to obtain analytic results due to the form of its Fourier transform and will therefore be the main focus of this paper. Note that this kernel further allows a transformation from a nonlocal to a local model by introducing an additional variable \cite{Britton1990, Gourley2001, Merchant2015}, but in the interest of considering other dispersal kernels we will not use this approach. To investigate the effects of the kind of decay of the kernel, we will finally consider the power law distribution
\begin{align}\label{Intro:power law distribution kernel}
\phi(x) = \frac{(b-1)a_p}{2\left(1+a_p|x|\right)^b}, \quad b >3,
\end{align}
where $x\in\R$, and $a_p >0$, $b>0$ are the scale and shape parameters of the distribution, respectively. Note that for this kernel function the derivation of the local model above does not hold. For a review of other biologically relevant plant dispersal kernels see \cite[Table 1]{Bullock2017}.

The purpose of this paper is to gain an understanding of how the shape of the dispersal kernel in the nonlocal model \eqref{eq:Intro Klausmeier nonlocal} affects the tendency to form patterns. In particular, we will mainly focus on the maximum rainfall parameter $A_{\max}$ that supports the formation of patterns, or in other words, the lowest amount of precipitation that allows plants to form a homogeneous vegetation cover. This critical rainfall level will be determined using different approaches for the Laplacian kernel \eqref{eq:Intro Laplacian kernel}. While all those approaches provide the same information on $A_{\max}$, they all give different further insights into other properties of the model. In Section \ref{sec:Linear Stability} we will investigate the model using linear stability analysis, obtaining information on the pattern wavelength alongside the upper bound on the rainfall. The constant uphill migration of the plants suggests studying the system in its travelling wave form. This will be done in Section \ref{sec:TWS}, where the critical rainfall level can be deduced from the loci of a Hopf bifurcation. Finally, the asymptotic form of the model is studied in Section \ref{sec:Asymptotics of Model}. All these approaches make use of the size of the parameter $\nu$ by obtaining conditions to leading order in $\nu$ as $\nu \rightarrow \infty$. A comparison to other dispersal kernels is shown in Section \ref{sec:Numerics} using numerical simulations of the model. From these we will be able to deduce parametric trends on how the tendency to form patterns is affected by the width and the type of decay of the dispersal. Finally, we discuss our results from an ecological viewpoint in Section \ref{sec:Comparison}. Motivated by the discussion above, the analysis will be done in three different cases; the situation in which $C= 2/\sigma(a)^2$, which allows us to compare our results for the Laplacian kernel to the corresponding results for the local model obtained by \cite{Sherratt2005, Sherratt2010, Sherratt2011, Sherratt2013III, Sherratt2013IV, Sherratt2013V}, and the cases in which one of $C$ or $a$ is kept constant, while the other parameter is varied. 


\section{Linear Stability Analysis}\label{sec:Linear Stability}
In this section we will use linear stability analysis to investigate to occurrence of spatial patterns in the nonlocal Klausmeier model \eqref{eq:Intro Klausmeier nonlocal} with the Laplacian kernel \eqref{eq:Intro Laplacian kernel}. We will show that the maximum rainfall parameter $A_{\max}$ supporting pattern formation is $O_s(\nu^{1/2})$ ($f=O_s(\nu) \Longleftrightarrow f = O(\nu)$ and $f \ne o(\nu)$), and will obtain an explicit expression for it. This will show that both an increase in $a$ for $C$ being kept constant and an increase in $C$ for $a$ being kept constant yields an increase of $A_{\max}$, while under the assumption that $C=a^2$ an increase in $a$ (and thus $C$) results in a decrease of the critical value $A_{\max}$. Further this analysis will allow us to investigate the wavelength of the patterned solutions of the model.


The steady states of \eqref{eq:Intro Klausmeier nonlocal} are
\begin{align*}
(\overline{u}_1,\overline{w}_1) &= (0,A), \quad (\overline{u}_2,\overline{w}_2) = \left(\frac{2B}{A-\sqrt{A^2-4B^2}},\frac{A-\sqrt{A^2-4B^2}}{2}\right), \\ (\overline{u}_3,\overline{w}_3) &= \left(\frac{2B}{A+\sqrt{A^2-4B^2}},\frac{A+\sqrt{A^2-4B^2}}{2}\right),
\end{align*}
where $(\overline{u}_2,\overline{w}_2)$ and $(\overline{u}_3, \overline{w}_3)$ only exist if $A\ge2B$. The steady state $(\overline{u}_1,\overline{w}_1)$ describing extinction of plants $u$ is always stable, while $(\overline{u}_3,\overline{w}_3)$ is unstable for all choices of parameters, provided it exists. The steady state $(\overline{u},\overline{w}):=(\overline{u}_2,\overline{w}_2)$ is stable to spatially homogeneous perturbations if $B<2$. For $B>2$, it is only stable for sufficiently large values of $A$. Estimates of the parameters, however, suggest that $B<2$.

To investigate the possibility of spatial patterns, consider spatially heterogeneous perturbations $u=\overline{u}+\tilde{u}(x,t)$, $w= \overline{w}+\tilde{w}(x,t)$ proportional to $e^{\lambda t+ikx}$ for growth rate $\lambda \in \C$ and wavenumber $k>0$. Linearising the resulting system gives that $\lambda$ satisfies the dispersion relation
\begin{align*}
\lambda = \frac{1}{2} \left( C\left(\widehat{\phi}(k)-1\right)-dk^2+\alpha+\delta+i\nu k \pm \sqrt{R+iI} \right),
\end{align*}
where $\widehat{\phi}(k)$ is the Fourier transform of $\phi$,
\begin{equation*}
\begin{aligned}
R &= \left(C\left(\widehat{\phi}(k)-1 \right)+dk^2 \right)^2 +2C\left(\widehat{\phi}(k)-1 \right)(\alpha-\delta) \\ &+\left(2\alpha d-2\delta d -\nu^2 \right)k^2+4\gamma\beta +(\alpha-\delta)^2,   
\end{aligned}
\end{equation*}
and
\begin{align*}
I &= -2\nu k\left(C\left(\widehat{\phi}(k)-1 \right) +dk^2+\alpha-\delta \right).
\end{align*}
For a Turing-Hopf bifurcation to occur, at least one eigenvalue needs to have positive real part. Therefore, the condition for patterns to form is 
\begin{align}\label{eq:real part lambda advection nonlocal}
\Re(\lambda) = \frac{1}{2}\left(\alpha + \delta - dk^2 +C\left(\widehat{\phi}(k)-1 \right) + \frac{1}{\sqrt{2}}\left(\sqrt{R^2+I^2}+R \right)^{\frac{1}{2}} \right) > 0.
\end{align}
To investigate this further for $\phi$ being the Laplacian kernel \eqref{eq:Intro Laplacian kernel}, we will make use of $\nu\gg 1$, by expanding \eqref{eq:real part lambda advection nonlocal} in $\nu$. With all other parameters $O_s(1)$ as $\nu \rightarrow \infty$, this gives
\begin{align}\label{eq:real part lambda advection nonlocal expanded with k=O(1)}
\Re(\lambda) = \alpha - \frac{Ck^2}{a^2+k^2}+ O\left(\frac{1}{\nu^2}\right),
\end{align}
provided that $(a^2+k^2)(\delta-\alpha-dk^2)+Ck^2<0$. If this condition is not satisfied, the expansion is $\Re(\lambda) = -dk^2+\delta <0$ for any $k>0$. Substituting $k=0$ into \eqref{eq:real part lambda advection nonlocal expanded with k=O(1)}, yields $\Re(\lambda) = \alpha>0$, which contradicts the stability of $(\overline{u},\overline{w})$ to spatially homogeneous perturbations. The occurrence of patterns is captured by assuming that $A$ is $O_s(\nu^{1/2})$. Expanding in $\nu \gg 1$ then gives
\begin{align}\label{eq:Linear Stability Analysis nonlocal advection expansion with scaling on A}
\Re(\lambda) = -{\frac { \left( -{B}^{5}{\nu}^{2}+{B}^{4}C{\nu}^{2} \right) {k}^{4}+
		\left( -{B}^{5}{a}^{2}{\nu}^{2}+{A}^{4}B+{A}^{4}C \right) {k}^{2}+{A}
		^{4}B{a}^{2}}{ \left( {B}^{4}{\nu}^{2}{k}^{2}+{A}^{4} \right)  \left( 
		{a}^{2}+{k}^{2} \right) }} +O\left(\frac{1}{\nu}\right).
\end{align}
Therefore, $\Re(\lambda)>0$ if 
\begin{align*}
q\left(k^2\right):=\left( -{B}^{5}{\nu}^{2}+{B}^{4}C{\nu}^{2} \right) {k}^{4}+
\left( -{B}^{5}{a}^{2}{\nu}^{2}+{A}^{4}B+{A}^{4}C \right) {k}^{2}+{A}
^{4}B{a}^{2} <0.
\end{align*}
This polynomial in $k^2$ attains its minimum
\begin{align}\label{eq:Linear Stability roots q}
q\left(k_{\min}^2\right) = {\frac {- \left( B+C \right) ^{2}{A}^{8}-2\,{B}^{5}{a}^{2}{\nu}^{
			2} \left( B-3C \right) {A}^{4}-{B}^{10}{a}^{4}{\nu}^{4}}{4{B}^{4}{\nu
		}^{2} \left( B-C \right) }}, 
\end{align}
at
\begin{align*}
k_{\min}^2 ={\frac {-{B}^{5}{a}^{2}{\nu}^{2}+{A}^{4}B+{A}^{4}C}{2{B}^{4}{\nu}^
		{2} \left( B-C \right) }}.
\end{align*}
Solving $q(k_\min^2)<0$ for $A^4$ gives $A_1^4<A^4<A_2^4$, where $A_1^4<A_2^4$ are the roots of \eqref{eq:Linear Stability roots q}. Substituting $A_1^4$ into $k_{\min}^2$ gives $k_{\min}^2<0$, which contradicts $k_{\min} \in \R$. Therefore, the sufficient condition for patterns to occur is 
\begin{align}\label{eq:Linear Stability Analysis condition on A nonlocal}
A < A_{\max} =  \left(\frac{3C-B-2\sqrt{2C}\sqrt{C-B}}{\left(B+C\right)^2} \right)^{\frac{1}{4}} a^{\frac{1}{2}} B^{\frac{5}{4}} \nu^{\frac{1}{2}},
\end{align}
valid to leading order in $\nu$ as $\nu \rightarrow \infty$. As expected, setting $C=a^2$ and taking the limit $a \rightarrow \infty$ yields the corresponding condition for the local model obtained by \cite{Sherratt2013IV}, which is

\begin{align}\label{eq:Linear Stability Local Advection scaling on A condition on A}
A< A_{\max} =  \left( \sqrt{2}-1 \right)^{\frac{1}{2}} B^{\frac{5}{4}} \nu^{\frac{1}{2}}.
\end{align}


\subsection{Wavelength}
It is of interest to investigate the wavelength of the patterns. While a rigorous analysis of this requires tools from nonlinear analysis, one can obtain some information about the wavelength from the results obtained in this section. For this we will assume that the patterns are dominated by the wavenumber giving the largest growth, that is the wavenumber $k_\max$ giving the maximum of $\Re(\lambda)$ given in \eqref{eq:Linear Stability Analysis nonlocal advection expansion with scaling on A}. Differentiating $\Re(\lambda)$ with respect to $k^2$ shows that it obtains its maximum at
\begin{align*}
k_\max^2 = -{\frac {{A}^{2}a \left( 2{A}^{2}{B}^{2}a\nu \left( B-\frac{C}{2} \right) +\sqrt {2BC} \left( -{B}^{4}a^2\nu^2+{A}^{4} \right)\right) }{-{B}^{6}C{a}^{2}{\nu}^{3}+2{A}^{4}{B}^{3}\nu}}.
\end{align*}
Therefore the wavelength $L$ is given by
\begin{multline}\label{eq:Comparison Wavelength nonlocal}
L = \frac{2\pi}{k_\max} \\ = 2\pi \left(\frac{-{B}^{6}C{a}^{2}{\nu}^{3}+2{A}^{4}{B}^{3}\nu}{{A}^{2}a \left( {A}^{2}{B}^{2}a\nu \left( 2B-C \right) +\sqrt {2}\sqrt {BC \left( -{B}^{2}a\nu+{A}^{2} \right) ^{2} \left( {B}^{2}a\nu+{A}^{2} \right) ^{2}} \right) }\right)^{\frac{1}{2}}.
\end{multline}

\begin{figure}
		\centering
		\subfloat[\label{fig:Wavelengths}]{\includegraphics[width=0.49\textwidth]{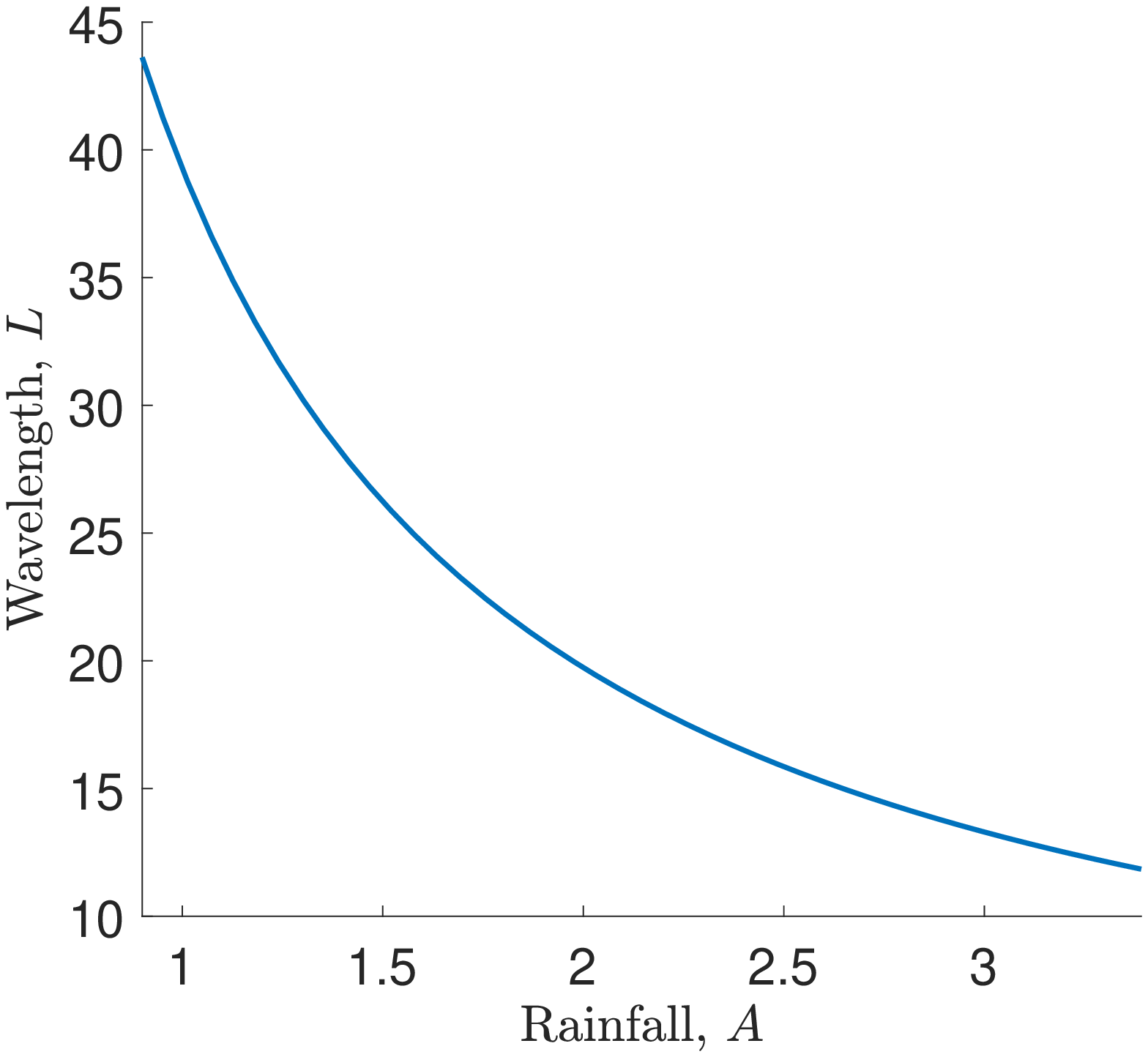}}
		\subfloat[\label{fig:Wavelength comparison}]{\includegraphics[width=0.49\textwidth]{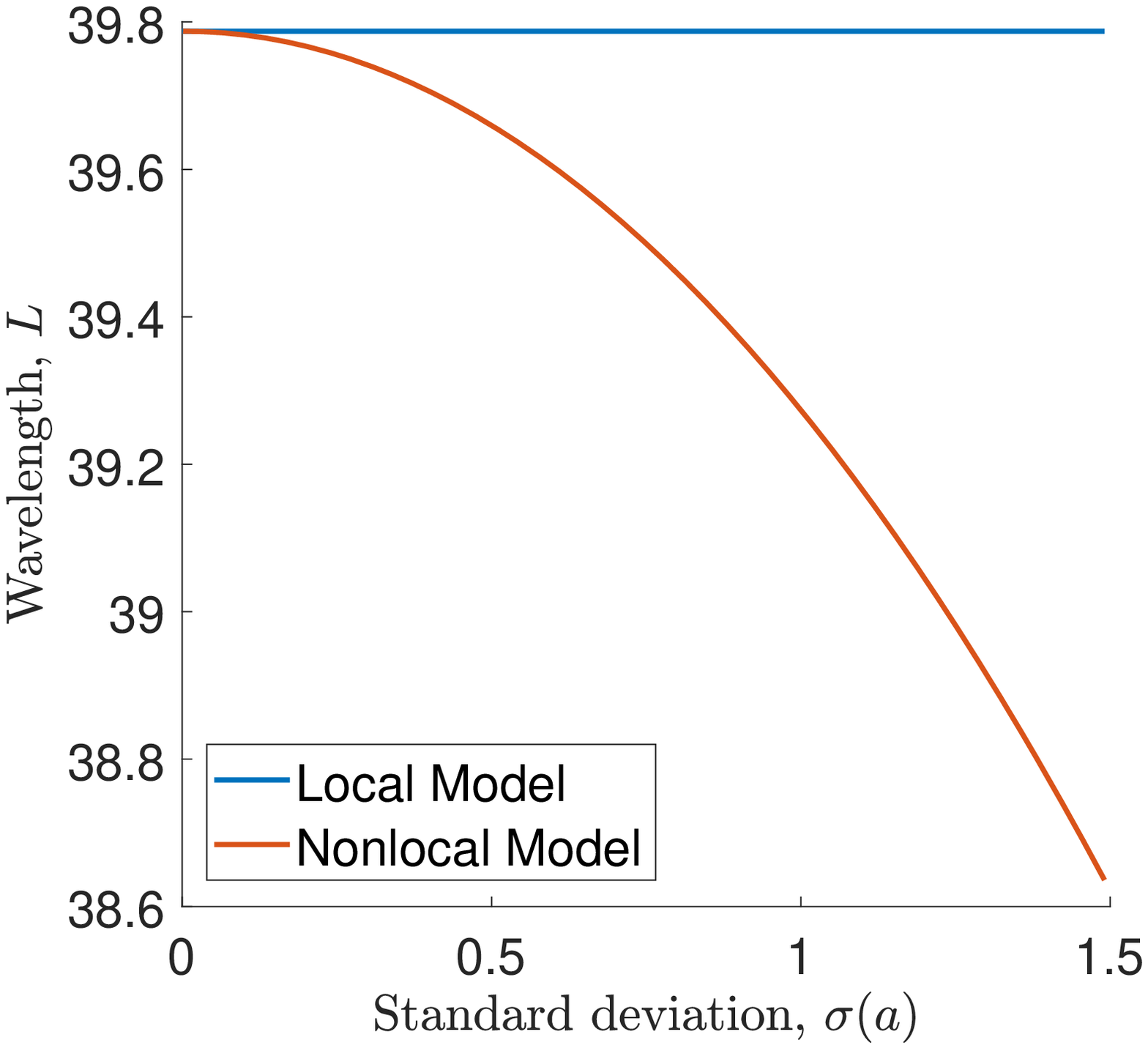}}
	\caption{Variation in pattern wavelength with rainfall $A$ and standard deviation $\sigma(a)$. Part (a) shows the wavelength \eqref{eq:Comparison Wavelength nonlocal} of the patterns as it decreases with the rainfall $A$ for nonlocal model. The parameter values are $B=0.45$, $\nu=182.5$, $C=1$ and $a=1$. Part (b) compares the wavelength predicted from the nonlocal model with the setting $C=a^2$ as it varies with the dispersal parameter $a$ and compares it to the wavelength obtained from the local model. It shows that the nonlocal model predicts a shorter distance between the vegetation stripes, especially if the shape of the dispersal kernel is wide. However, the difference is very small (see the $y$-axis of the plot). The parameter values used for this are $A=1$, $B=0.45$, $\nu=182.5$}
\end{figure}

The wavelength $L$ is decreasing in the rainfall parameter $A$, decreasing in the dispersal parameter $a$ if the dispersal coefficient $C$ is fixed, increasing in $C$ when $a$ is kept constant, and increasing in $a$ if one sets $C=a^2$. Figure \ref{fig:Wavelengths} shows the wavelength as it varies with the rainfall parameter $A$ for some fixed $B$, $C$, $a$ and $\nu$. Also note that when $C=a^2$, the wavelength \eqref{eq:Comparison Wavelength nonlocal} for the nonlocal model approaches the wavelength predicted by the local model as $a\rightarrow \infty$, as expected by the limiting behaviour of the nonlocal model. Combining these two results shows that the nonlocal model predicts a shorter distance between vegetation stripes than the local model with this setting of $C$. This is visualised in Figure \ref{fig:Wavelength comparison}.

\section{Travelling Wave Solutions}\label{sec:TWS}
The constant uphill migration of the vegetation patterns suggests considering travelling waves. In this section we will investigate the travelling wave form of the nonlocal Klausmeier model \eqref{eq:Intro Klausmeier nonlocal}. Pattern solutions of the original PDE model then correspond to periodic solutions of the travelling wave ODEs. From the equations in their travelling wave form we will not only be able to confirm the results on the maximum rainfall supporting pattern formation obtained by performing linear stability analysis in Section \ref{sec:Linear Stability}, but also deduce more information about the migration speed of the patterns. The nature of the patterned solutions fundamentally depends on the scaling of the migration speed $c$. The highest rainfall level supporting pattern formation occurs for $c=O_s(1)$. For this situation we determine conditions for Hopf bifurcations to occur; for the local Klausmeier model \eqref{eq:Intro Klausmeier local} the parameter range in the $A$-$c$ plane that supports pattern formation is bounded above by the locus of a Hopf bifurcation \cite{Sherratt2007}, and we anticipate the same for the nonlocal model.

Applying the travelling wave ansatz $u(x,t) = U(z)$, $w(x,t)=W(z)$, $z=x-ct$ to the nonlocal model \eqref{eq:Intro Klausmeier nonlocal}, gives
	\begin{align*}
	\frac{\dif U}{\dif z} &= -\frac{1}{c} \left(U^2W-BU + C\left(\int_{-\infty}^\infty \phi(z-z')U(z') \dif z' - U(z)\right) \right),  \\
	\frac{\dif W}{\dif z} &= -\frac{1}{c+\nu} \left(A-W-U^2W+d\frac{\dif^2 W}{\dif z^2} \right).
	\end{align*}
To investigate the occurrence of a Hopf bifurcation, consider perturbations $\tilde{U}(z)$, $\tilde{W}(z)$ proportional to $e^{\lambda z}$ of the steady state $(\overline{U},\overline{W}) = (\overline{u},\overline{w})$. Setting $\phi$ to be the Laplacian kernel \eqref{eq:Intro Laplacian kernel} and linearising the resulting system gives that $\lambda$ satisfies
\begin{align}\label{eq:Eigenvalues nonlocal TWS}
\lambda^5+\alpha\lambda^4+\beta\lambda^3+\gamma\lambda^2+\delta\lambda+\varepsilon = 0,
\end{align}
where
\begin{align*}
\alpha &= \frac{d(B-C)+c(c+\nu)} {cd}, \\ 
\beta &= \frac{-2B^2\left(a^2cd-(B-C)(c+\nu)  \right)-Ac\left(A+\sqrt{A^2-4B^2}\right)}{2B^2cd}, \\
\gamma &= \frac{-2B^2a^2 \left(d+c(c+\nu) \right) +A(B+C)\left(A+\sqrt{A^2-4B^2}\right)-4B^3}{2B^2cd}, \\ 
\delta &= \frac{a^2\left(-2B^3(c+\nu) +Ac\left(A+\sqrt{A^2-4B^2}\right)\right)}{2B^2cd}, \\ 
\varepsilon &= \frac{a^2\left(-A\left(A+\sqrt{A^2-4B^2}\right) + 4B^2 \right)}{2B^2cd}.
\end{align*}
To find conditions for a Hopf bifurcation to occur, set $\lambda=i\omega$, $\omega \in \R$. This splits \eqref{eq:Eigenvalues nonlocal TWS} into its real and imaginary parts, which after solving for and eliminating $\omega^2$ gives the condition
\begin{align}\label{eq:Hopf Condition TWS Nonlocal}
\frac{\gamma \pm \sqrt{\gamma^2-4\alpha\varepsilon}}{2\alpha} = \frac{\beta \pm \sqrt{\beta^2-4\delta}}{2}.
\end{align}
The assumption $\omega \in \R$ requires that the left and right hand sides of this equation are both positive. This leads to an additional condition \eqref{eq:TWS condition on c that omega real, d not 0} that will be considered later. To further investigate \eqref{eq:Hopf Condition TWS Nonlocal}, we expand it in $1/\nu$. This gives
\begin{align*}
\frac{((B-C)\operatorname{sign}(c)+B+C)a^2}{B-C} + O\left(\frac{1}{\nu}\right) = 0.
\end{align*}
For the first term of the expansion to be zero, one would require $B>C$ with one of the parameters being equal to zero, depending on the sign of $c$. This is, however, not possible due to the positivity assumptions on the parameters. Investigating the next term of the expansion suggests using the scaling $A=O_s(\nu^{1/2})$. Applying this scaling to \eqref{eq:Hopf Condition TWS Nonlocal}, expanding in $\nu\gg 1$ and then solving for $c$ shows that a Hopf bifurcation exists at
\begin{multline}\label{eq:Hopf condition on c Nonlocal d not 0}
c_\pm = \left(\frac{B}{2A^2} +\frac{A^2(2B-C)}{2\left(-B^4a^2\nu^2+A^4 \right)}\right. \\ \pm \left. \left(\frac{B^2}{4A^4} + \frac{3BC}{2\left(-B^4a^2\nu^2+A^4 \right)} +\frac{4B^6a^2\nu^2 - 4A^4BC +A^4C^2}{4\left(-B^4a^2\nu^2+A^4 \right)^2} \right)^{\frac{1}{2}} \right)\nu B^2,
\end{multline}
to leading order in $\nu$ as $\nu \rightarrow \infty$. Since the migration speed $c \in \R$, this requires
\begin{align}\label{eq:Hopf condition on A Nonlocal d not 0}
A < A_{\max} = \left(\frac{3C-B-2\sqrt{2C}\sqrt{C-B}}{\left(B+C\right)^2} \right)^{\frac{1}{4}} a^{\frac{1}{2}} B^{\frac{5}{4}} \nu^{\frac{1}{2}},
\end{align}
and $C>B$. This is the same condition as \eqref{eq:Linear Stability Analysis condition on A nonlocal} obtained in Section \ref{sec:Linear Stability}. 

In deriving this condition we assumed that the terms in \eqref{eq:Hopf Condition TWS Nonlocal} were positive. By applying the scaling $A=O_s(\nu^{1/2})$ and expanding in $\nu \gg 1$, this yields the bounds
\begin{align}\label{eq:TWS condition on c that omega real, d not 0}
\max\left\{0, \frac{B^2(B-C)\nu}{A^2} \right\} < c < \frac{B^3 \nu}{A^2},
\end{align}
to leading order in $\nu$. This condition is satisfied if $C>B$. In the case of $C=a^2$ it holds if $a>\sqrt{B}$.

\begin{figure}
	\centering
	\subfloat[\label{fig:Travelling Wave A condition comparison}]{\includegraphics[width=0.49\textwidth]{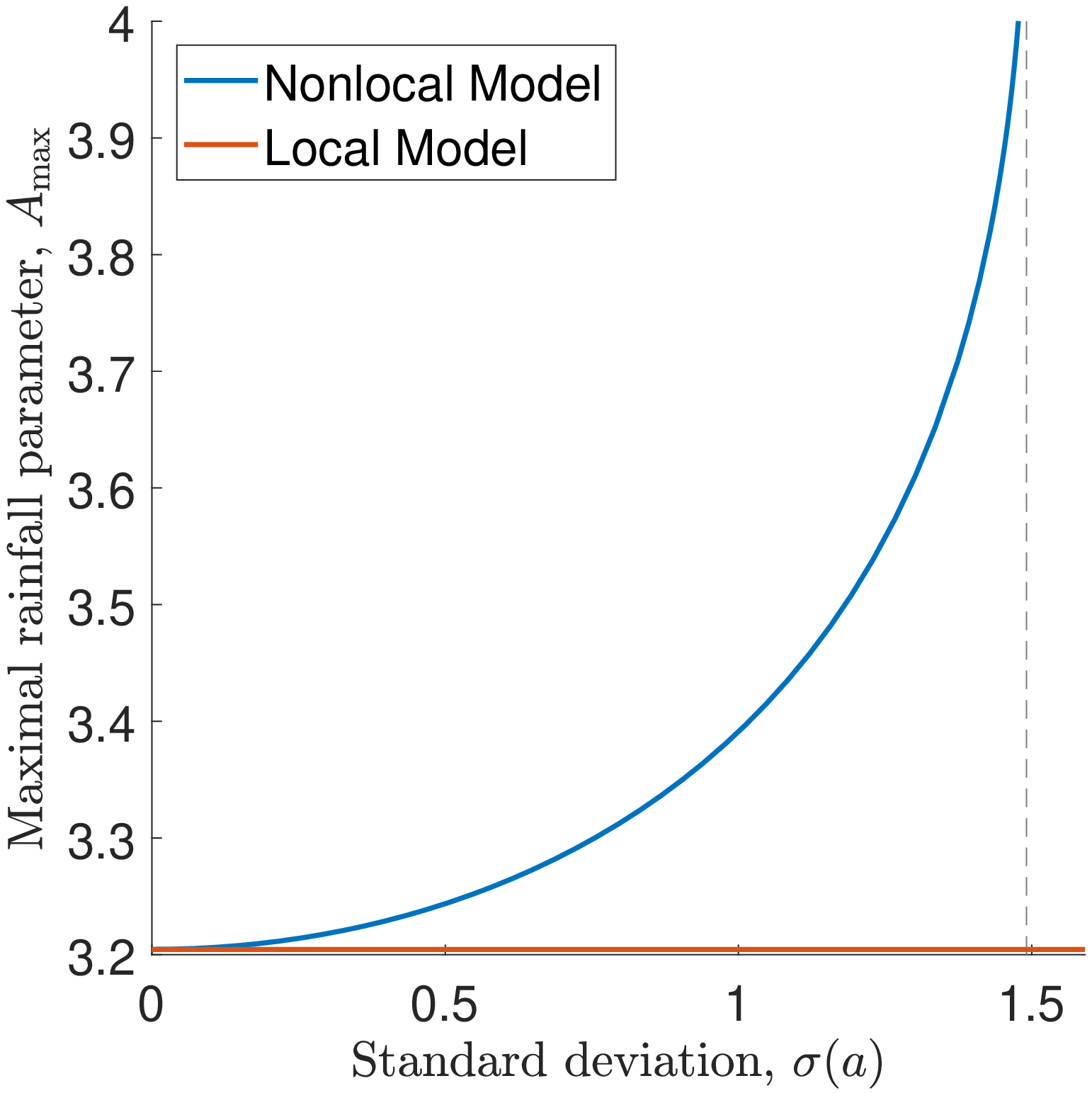}}
	\subfloat[\label{fig:TWS Ac plane nonlocal comparison}]{\includegraphics[width=0.49\textwidth]{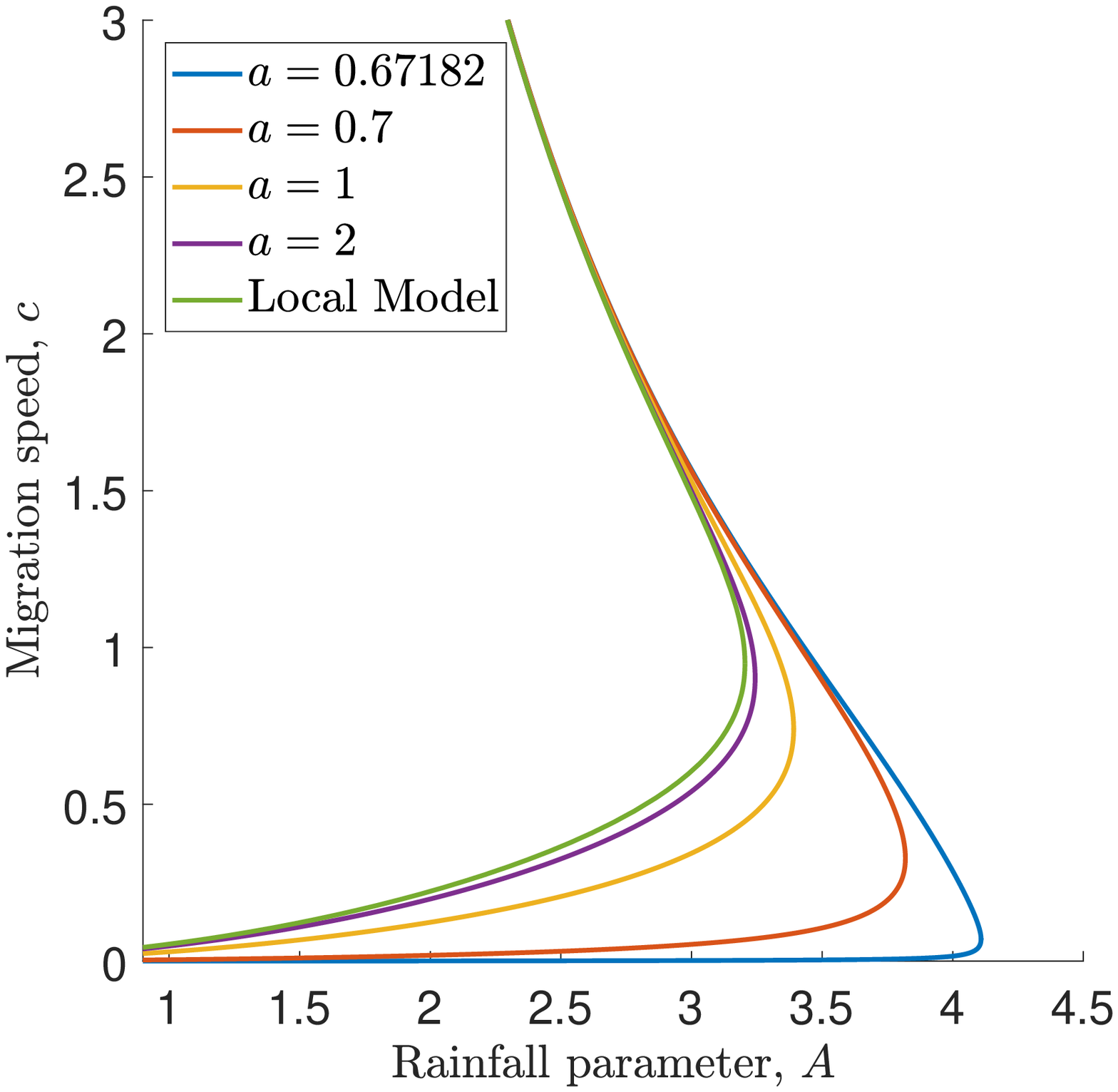}}
	\caption{Variation in the loci of the Hopf bifurcation and maximum rainfall parameter $A_{\max}$ with kernel width in the case $C=a^2$. The plot in (a) compares the upper bound \eqref{eq:Linear Stability Analysis condition on A nonlocal} on the rainfall parameter $A$ of the nonlocal model using the Laplacian kernel with $C=a^2$ with condition \eqref{eq:Linear Stability Local Advection scaling on A condition on A} obtained for the local model. Note that one requires $a>\sqrt{B}$ for $A_\max \in \R$ in the case of the nonlocal model.
		Part (b) compares the loci \eqref{eq:Hopf condition on c Nonlocal d not 0} of the Hopf bifurcations of the nonlocal model for different values of the dispersal parameter $a$ to the locus of the local model obtained by \cite{Sherratt2013IV}. The parameter values used in both figures are $B=0.45$, $\nu=182.5$}
\end{figure}

Setting $C=a^2$ and taking the limit $a\rightarrow \infty$ in both \eqref{eq:Hopf condition on c Nonlocal d not 0} and \eqref{eq:Hopf condition on A Nonlocal d not 0} gives, as expected by the considerations on the limiting behaviour of the model, the corresponding conditions obtained by \cite{Sherratt2013IV} for the local model. Further the right hand side of \eqref{eq:Hopf condition on A Nonlocal d not 0} is decreasing for all $a>\sqrt{B}$ in the setting $C=a^2$. Combined with the observation that it approaches the corresponding condition for the local model as $a \rightarrow \infty$, this shows that pattern formation is more likely in the nonlocal model with the tendency to form patterns increasing as the dispersal parameter $a$ decreases, i.e. as the width of the kernel $\phi$ increases. Figure \ref{fig:Travelling Wave A condition comparison} shows this for some fixed parameter values.
Finally, Figure \ref{fig:TWS Ac plane nonlocal comparison} combines these considerations by showing the loci \eqref{eq:Hopf condition on c Nonlocal d not 0} of the Hopf bifurcations of the nonlocal model for different values of the dispersal parameter $a$ in the $A$-$c$ plane and compares it to the corresponding locus of the local model. As shown previously, this implies that in the nonlocal model a larger parameter region supports pattern formation, especially as the dispersal parameter $a$ is decreased, i.e. as the width of the kernel is increased. This means that the nonlocal model predicts that plants which disperse their seeds over a larger distance will undergo a change from homogeneous vegetation to patterns at a higher level of rainfall than those plants with a narrower and diffusion-like dispersal, as the amount of rainfall is gradually decreased.

\begin{figure}
	\centering
	\subfloat[\label{fig:Comparison change Amax fixed C}]{\includegraphics[width=0.49\textwidth]{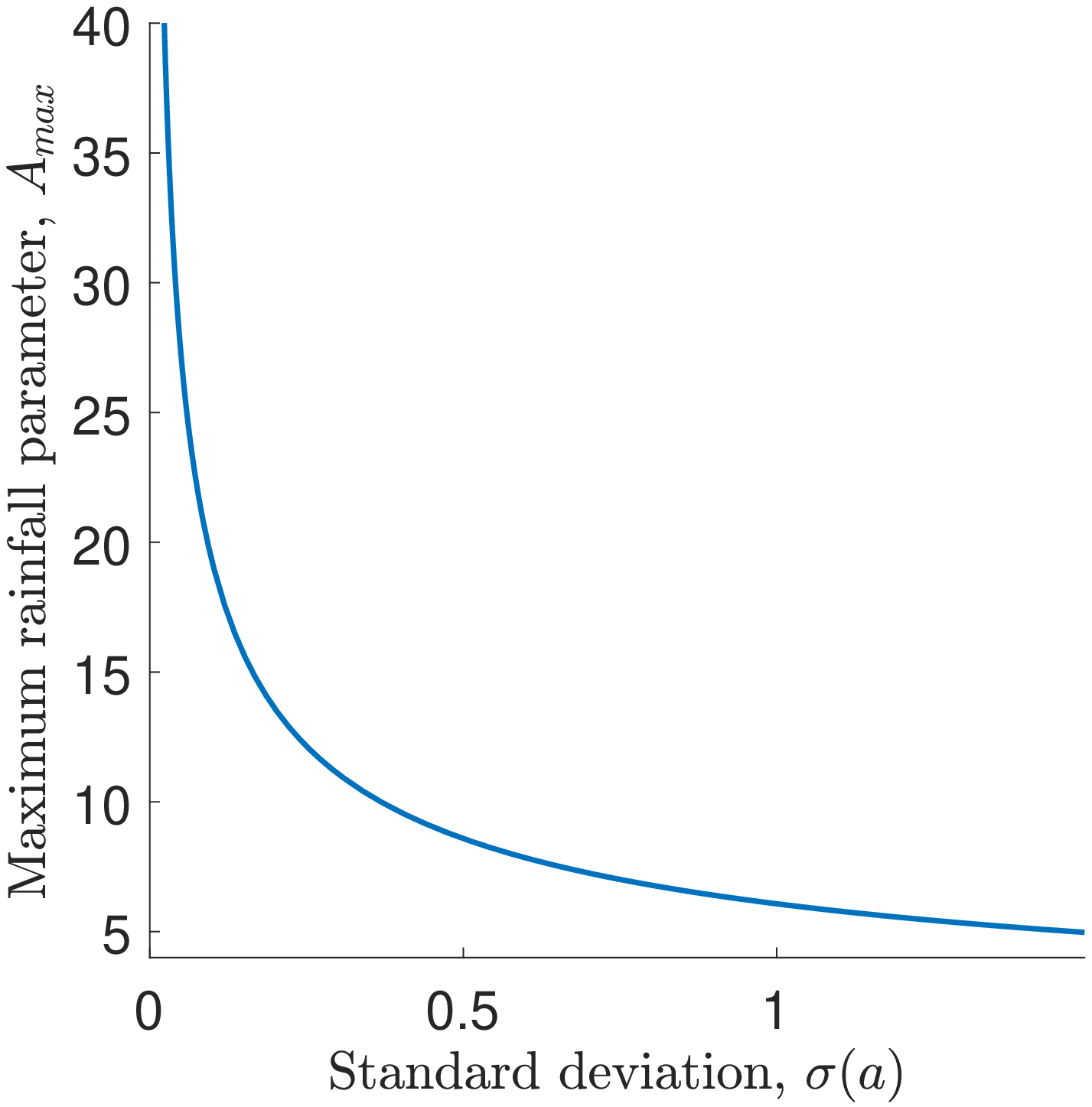}}		
	\subfloat[\label{fig:TWS Ac plane nonlocal C fixed}]{\includegraphics[width=0.49\textwidth]{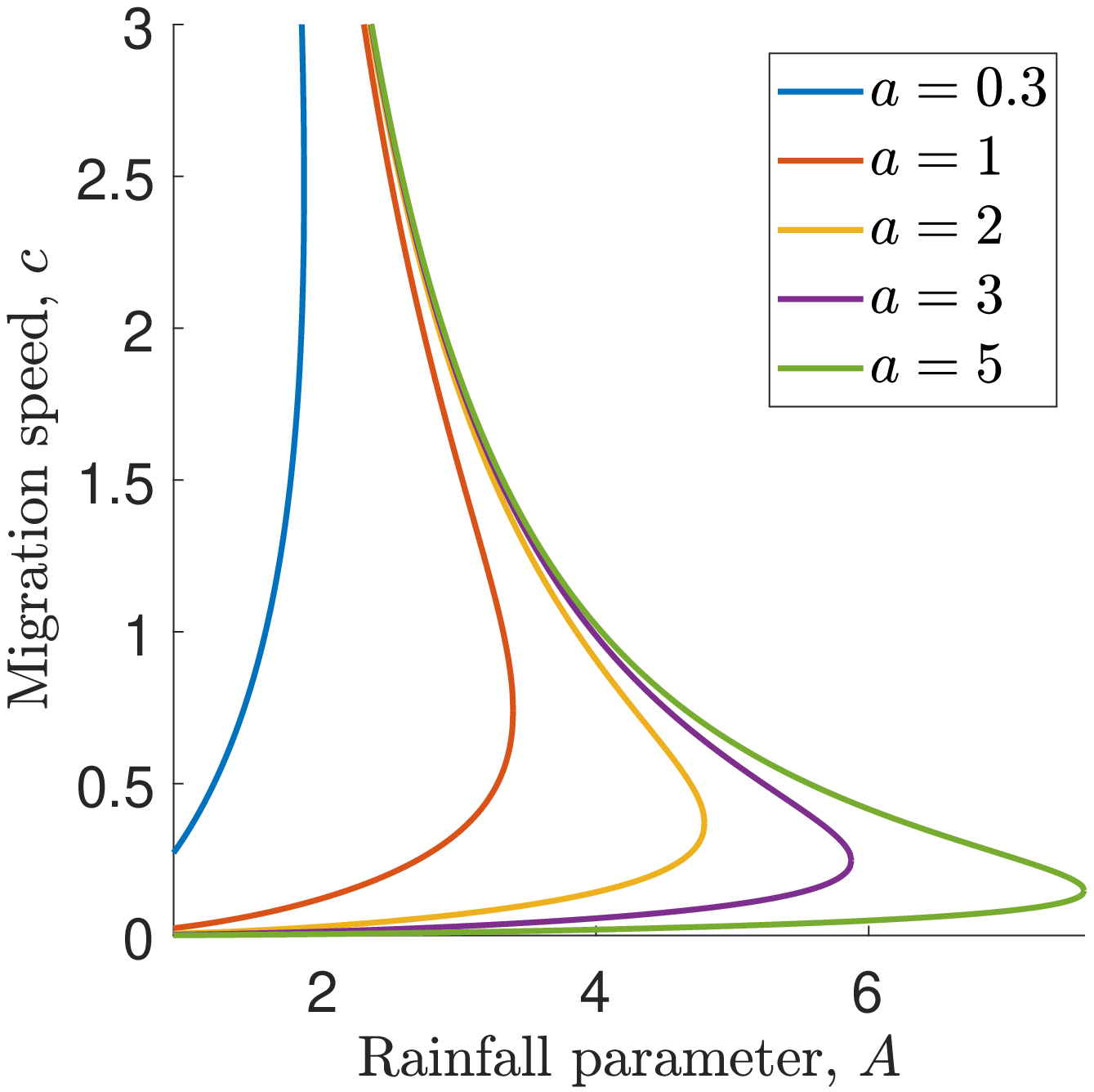}}
	\caption{Variation in the loci of the Hopf bifurcation and maximum rainfall parameter $A_{\max}$ with kernel width in the case of constant $C$. The plot in (a) shows how upper bound $A_\max$ given in \eqref{eq:Hopf condition on A Nonlocal d not 0} of the rainfall parameter $A$ that supports pattern formation in the nonlocal model using the Laplacian kernel varies as the dispersal parameter $a$ is changed. Here $C=1$ is fixed. Part (b) shows the loci \eqref{eq:Hopf condition on c Nonlocal d not 0} of the Hopf bifurcations of the nonlocal model with the Laplacian kernel for different values of the dispersal parameter $a$, where the dispersal coefficient $C$ is constant. The parameter values used here are $B=0.45$, $C=1$, $\nu=182.5$}
\end{figure}

If $C\ne a^2$, it is not appropriate to compare the nonlocal model to the local model. However, one can still investigate how a change in the dispersal parameter $a$ affects the tendency to form patterns in this situation. We will first consider the case in which $C$ is constant. In this situation \eqref{eq:Hopf condition on A Nonlocal d not 0} yields that the highest rainfall parameter supporting pattern formation $A_{\max}$ is proportional to $a^{1/2}$. This means that if the dispersal kernel gets narrower, a larger range of the rainfall parameter $A$ supports pattern formation. This is visualised in Figure \ref{fig:Comparison change Amax fixed C}, which shows the maximum rainfall parameter $A_{\max}$ plotted against the dispersal parameter $a$ and in Figure \ref{fig:TWS Ac plane nonlocal C fixed}, which visualises the location of the Hopf bifurcation \eqref{eq:Hopf condition on c Nonlocal d not 0}, where $C$ is constant. This is contrary to the behaviour observed in the case of $C=a^2$, where a narrower kernel gave less tendency to form patterns.

\begin{figure}
	\centering
	\subfloat[\label{fig:Comparison change Amax fixed a}]{\includegraphics[width=0.49\textwidth]{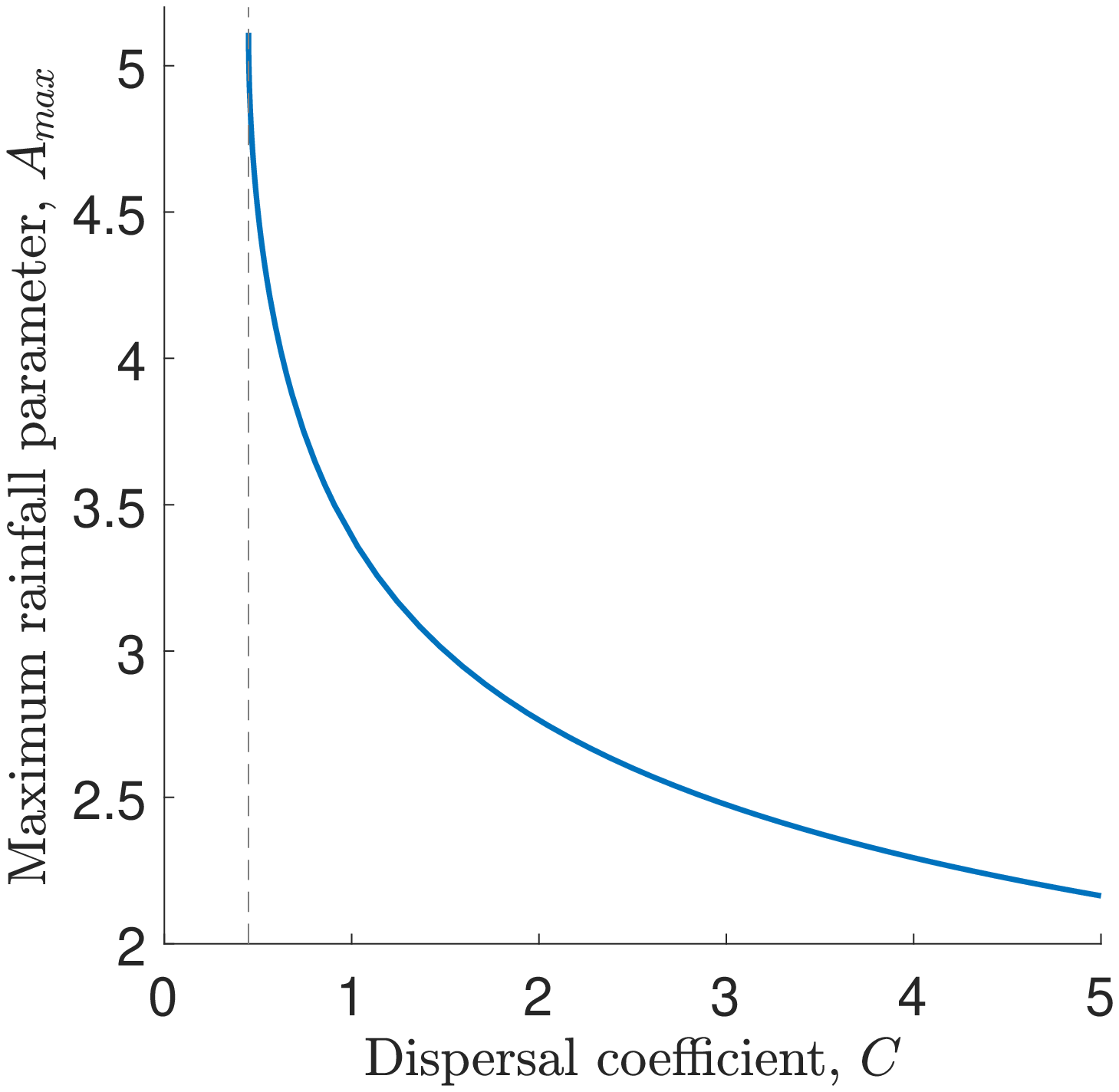}}
	\subfloat[\label{fig:TWS Ac plane nonlocal a fixed}]{\includegraphics[width=0.49\textwidth]{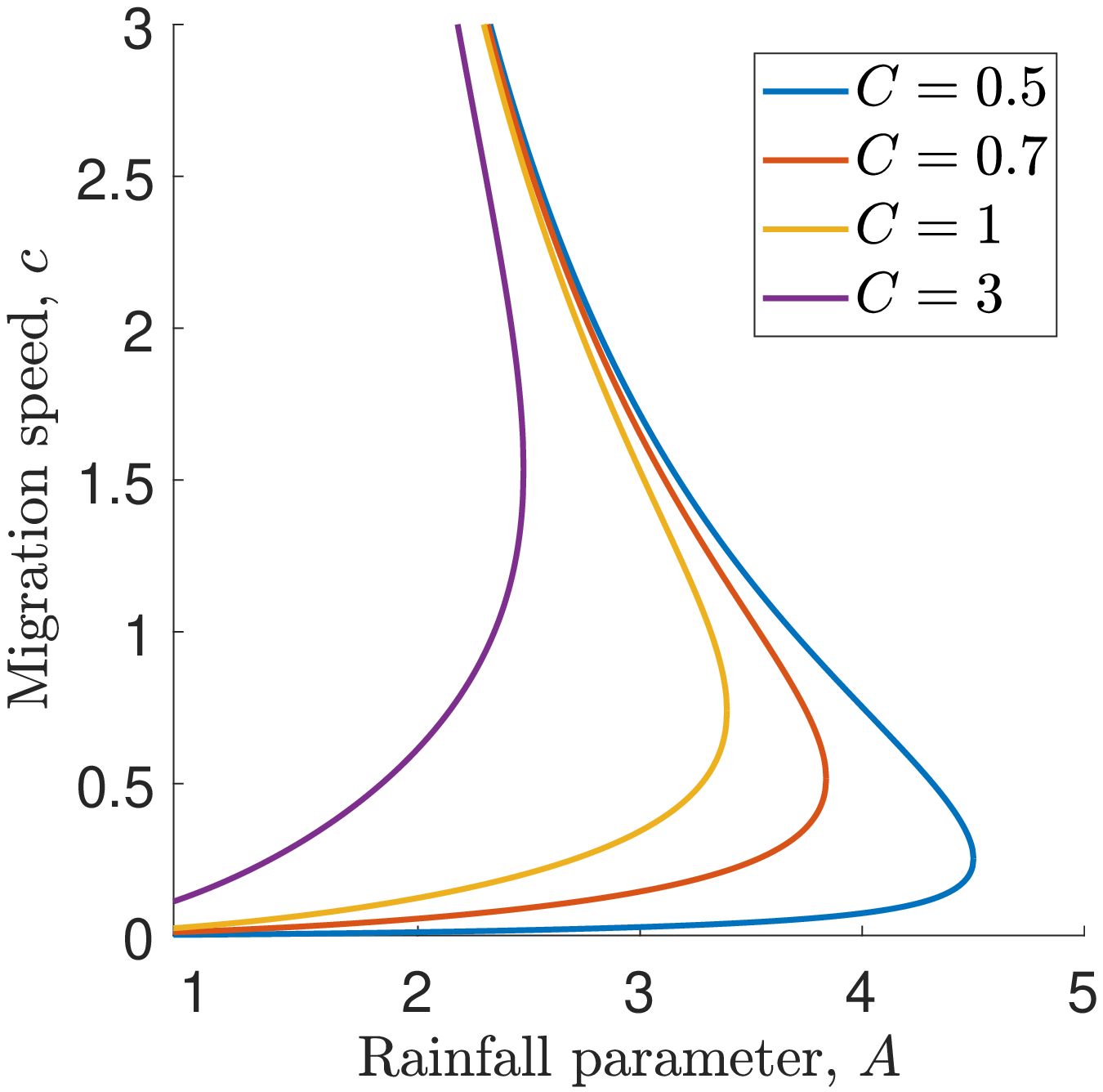}}
	\caption{Variation in the loci of the Hopf bifurcation and maximum rainfall parameter $A_{\max}$ with kernel width in the case of constant $a$. The plot in (a) shows how upper bound $A_\max$ given in \eqref{eq:Hopf condition on A Nonlocal d not 0} of the rainfall parameter $A$ that supports pattern formation in the nonlocal model with Laplacian kernel varies as the dispersal coefficient $C$ is changed. Here $a=1$ is fixed. Note that $C>B$ is required for $A_{\max} \in \R$. Part (b)  shows the loci \eqref{eq:Hopf condition on c Nonlocal d not 0} of the Hopf bifurcations in the same situation. The parameter values used here are $B=0.45$, $C=1$, $\nu=182.5$}
\end{figure}

Investigating the final case, i.e. the one of fixed $a$ and varying $C$, shows that the critical rainfall parameter $A_{\max}$ is decreasing with increasing $C$ for all $C>B$. This shows that the more the plants invest in their dispersal, the less likely is the formation of patterns. Similar to the previous two cases, the change in $A_{\max}$ is visualised in Figure \ref{fig:Comparison change Amax fixed a} and the loci of the Hopf bifurcations in the $A$-$c$ plane is shown in Figure \ref{fig:TWS Ac plane nonlocal a fixed}.

\section{Asymptotic Analysis of the Integro-PDE Model}\label{sec:Asymptotics of Model}
In the previous sections we have applied different techniques to the model \eqref{eq:Intro Klausmeier nonlocal} to find conditions for pattern formation in their leading order form. In this section we will confirm these by first obtaining the leading order form of the Integro-PDE model and then deducing conditions for Hopf bifurcations from it.

Applying the rescalings $u=AB^{-1}u^*$, $w=A^{-1}B^2w^*$, $t=B^{-1}t^*$, $c=Bc^*$, $\nu = A^2B^{-2}\Gamma^{-1}$, $B^{-1}C = D$ to  \eqref{eq:Intro Klausmeier nonlocal} gives
\begin{align*}
\frac{\partial u}{\partial t} &= u^2w - u + D\left(\int_{-\infty}^\infty \phi(x-y)u(y,t) \dif y - u(x,t)\right),  \\
B\nu^{-1}\frac{\partial w}{\partial t} &=\Gamma \left( 1-u^2w\right) - \nu^{-1}w + \frac{\partial w}{\partial x}+ d\nu^{-1} \frac{\partial^2w}{\partial x^2},
\end{align*}
where the $^*$'s were dropped for brevity. Again assuming that $A=O_s(\nu^{1/2})$, the leading order form in $\nu$ of this is
	\begin{align*}
	\frac{\partial u}{\partial t} &= u^2w - u + D\left(\int_{-\infty}^\infty \phi(x-y)u(y,t) \dif y - u(x,t)\right),  \\
	0 &=\Gamma \left( 1-u^2w\right)  + \frac{\partial w}{\partial x}.
	\end{align*}
Applying the travelling wave ansatz $u(x,t) = U(z)$, $w(x,t)=W(z)$, $z=x-ct$, gives
	\begin{align*}
	-c\frac{\dif U}{\dif z} &=  U^2W-U + D\left(\int_{-\infty}^\infty \phi(z-z')U(z') \dif z' - U(z)\right) ,  \\
	0 &= \Gamma\left(1-U^2W\right) + \frac{\dif W}{\dif z}.
	\end{align*}
This system has a unique steady state given by $(\overline{U},\overline{W}) = (1,1)$. Consider small perturbations $\tilde{U}$, $\tilde{W}$ of the steady state that are proportional to $e^{\lambda z}$. Letting $\phi$ to be the Laplacian kernel \eqref{eq:Intro Laplacian kernel} and linearising the resulting system yields that $\lambda$ satisfies
\begin{align}\label{eq:Asymptotic Model dispersion relation Laplacian}
\lambda^4 +\overline{\alpha}\lambda^3 + \overline{\beta}\lambda^2 + \overline{\gamma}\lambda + \overline{\delta} = 0,
\end{align}
where
\begin{align*}
\overline{\alpha} = \frac{1-D-\Gamma c}{c}, \quad \overline{\beta}=\frac{\Gamma(1-D) - a^2c}{c}, \quad \overline{\gamma} = \frac{a^2(\Gamma c -1)}{c}, \quad \overline{\delta} = -\frac{\Gamma a^2}{c}.
\end{align*}
To find conditions for a Hopf bifurcation to occur, again set $\lambda=i\omega$, $\omega \in \R$. Analogous to the preceding sections, this allows splitting \eqref{eq:Asymptotic Model dispersion relation Laplacian} into its real and imaginary parts, which after solving for $\omega^2$, assuming that $\omega\ne 0$, gives
\begin{align}\label{eq:Asymptotic Model omega squared}
\frac{\overline{\beta} \pm \sqrt{\overline{\beta}^2 - 4 \overline{\delta}}}{2} = \frac{\overline{\gamma}}{\overline{\alpha}},
\end{align}
as the leading order condition for a Hopf bifurcation to occur. The restriction $\omega \in \R$, implies the additional requirement
\begin{align}\label{eq:Asymptotic Model Nonlocal minor condition on c}
\max\left\{0, \frac{1-D}{\Gamma} \right\} < c < \frac{1}{\Gamma}.
\end{align} 
If $c>0$, then $\overline{\delta} <0$ and thus one needs to choose the plus sign on the left hand side in \eqref{eq:Asymptotic Model omega squared}. Solving for $c$  gives
\begin{align}\label{eq:Asymptotic Model Hopf condition c Gamma nonlocal}
c_\pm = {\frac { \left( D-3 \right) {\Gamma}^{2}+{a}^{2} \pm \sqrt { \left( D+1 \right) ^{2}{\Gamma}^{4}-2{a}^{2} \left( 3D-1 \right) {\Gamma}^{2}+{a}^{4}}}{2{\Gamma}\left( {a}^{2}-\Gamma^2\right)}}.
\end{align}
To satisfy \eqref{eq:Asymptotic Model Nonlocal minor condition on c}, one requires $C>B$ and 
\begin{align}\label{eq:Asymptotic Model Hopf condition Gamma nonlocal}
\Gamma < \left(\frac{3D-1+2\sqrt{2D(D-1)}}{(D+1)^2} \right)^{\frac{1}{2}}a.
\end{align}
Therefore, the steady state $(\overline{U}, \overline{W}) = (1,1)$ undergoes a Hopf bifurcation if \eqref{eq:Asymptotic Model Nonlocal minor condition on c}, \eqref{eq:Asymptotic Model Hopf condition c Gamma nonlocal} and \eqref{eq:Asymptotic Model Hopf condition Gamma nonlocal} are satisfied. Substituting the rescalings used at the beginning of this section into these three conditions gives the same conditions \eqref{eq:TWS condition on c that omega real, d not 0}, \eqref{eq:Hopf condition on c Nonlocal d not 0} and \eqref{eq:Hopf condition on A Nonlocal d not 0} that were obtained form the travelling wave equations in Section \ref{sec:TWS}.


\section{Numerical Simulations}\label{sec:Numerics}

So far, we have only considered one particular form of dispersal kernel in the nonlocal Klausmeier model \eqref{eq:Intro Klausmeier nonlocal}. In this section we will solve the model numerically for different kernel functions and use the solutions to estimate the maximum rainfall parameter giving patterns for each kernel. The simulations will show that the parametric trends that were obtained for the Laplacian kernel carry over to other kernel functions, i.e. a wider dispersal kernel and a higher dispersal rate decrease the tendency to form patterns, while under the assumption that $C=2/\sigma(a)^2$, an increase in kernel width causes an increase in the size of the parameter region giving patterns. Our numerical simulations will further show that the tendency to form patterns depends on the type of decay of the dispersal kernel.

\begin{figure}
	\centering
	\subfloat[\label{fig:Numerics nonlocal solution no patterns}]{\includegraphics[width=0.9\textwidth]{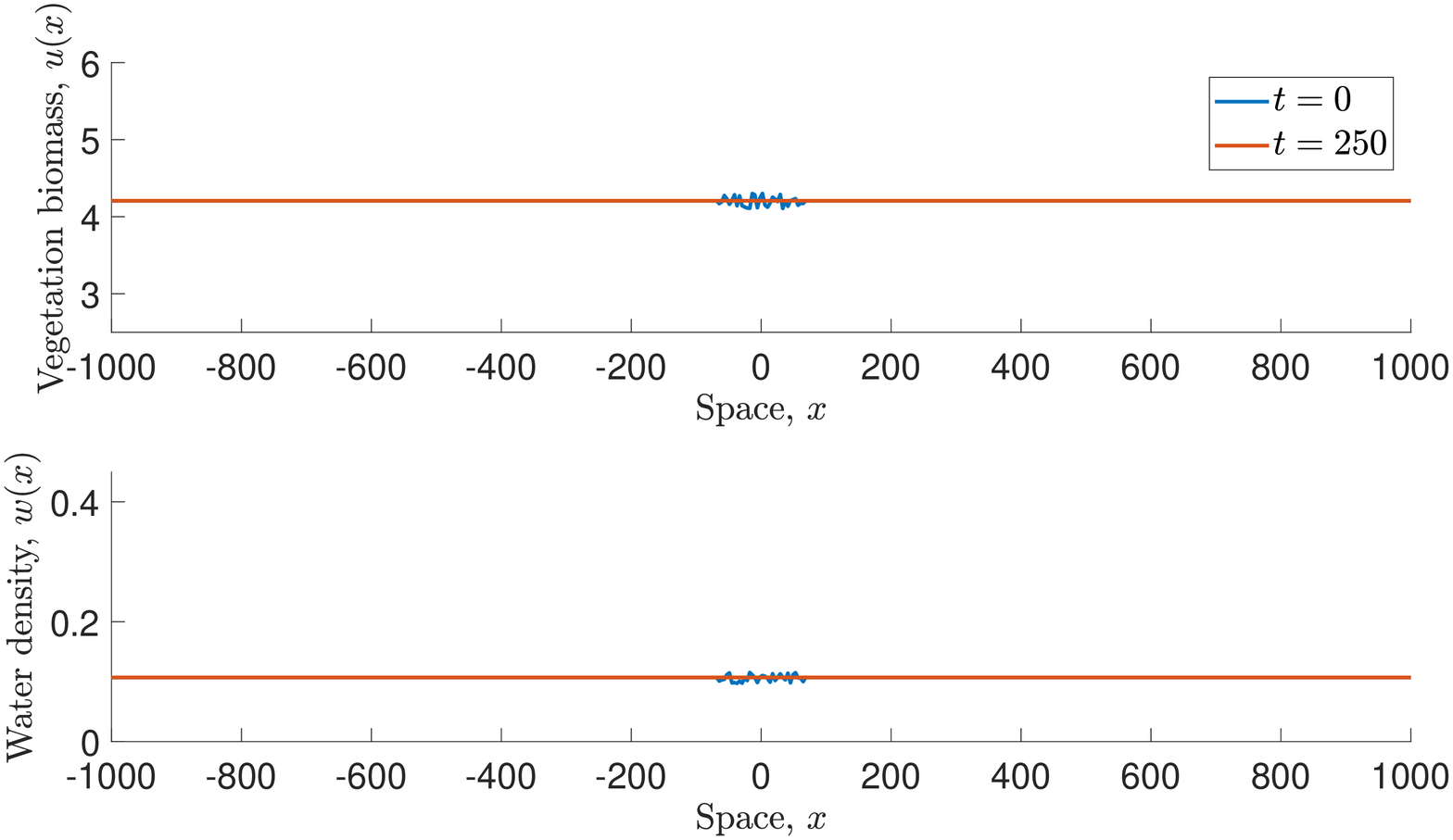}} \\
	\subfloat[\label{fig:Numerics nonlocal solution whole interval}]{\includegraphics[width=0.9\textwidth]{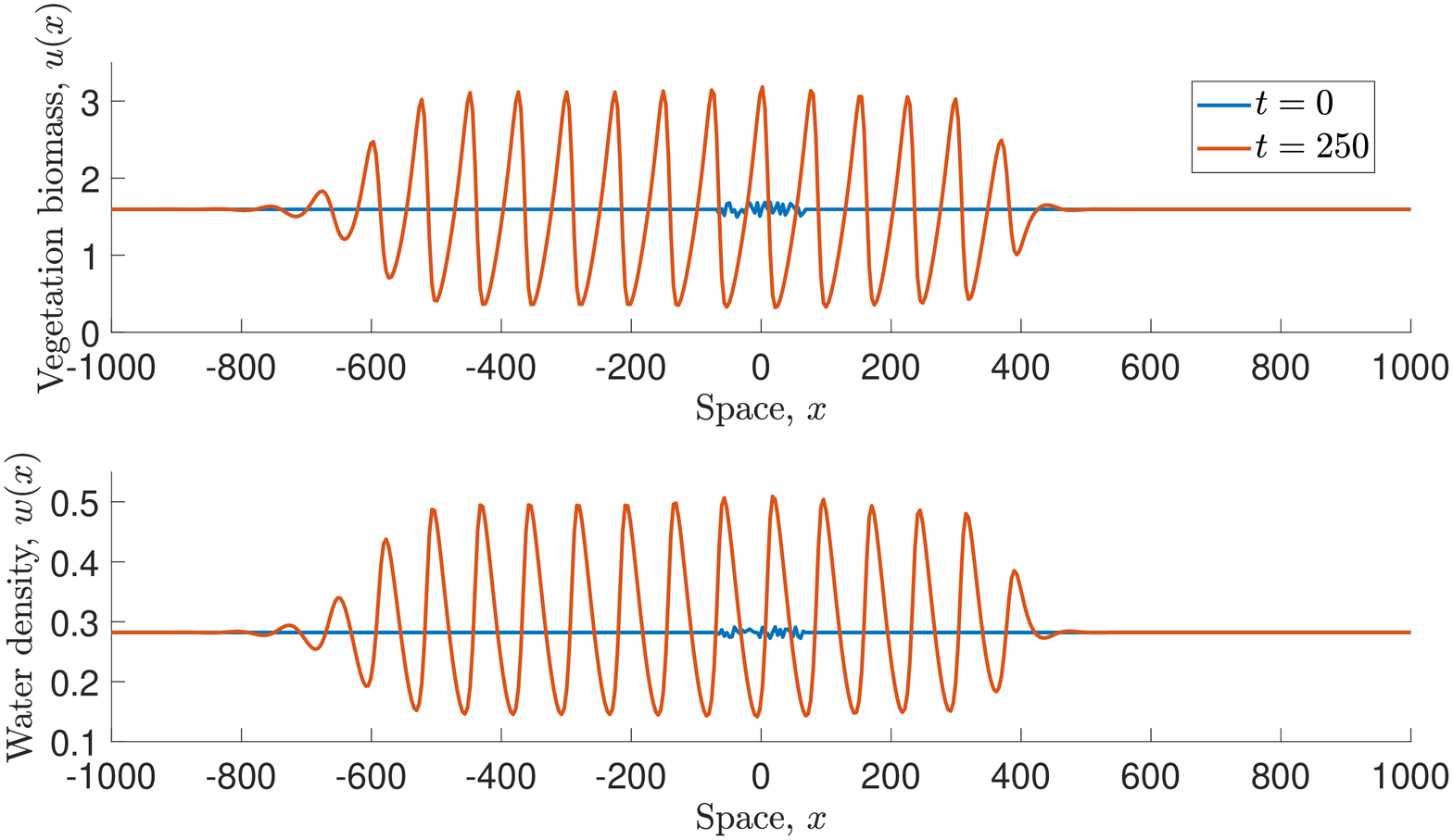}}
	\caption{Numerical solution of the nonlocal Klausmeier model \eqref{eq:Intro Klausmeier nonlocal} using the Laplacian kernel \eqref{eq:Intro Laplacian kernel} for different rainfall levels. In (a) $A=2$ yields convergence to the coexisting steady state from which the system is perturbed initially. Part (b) displays a patterned solution obtained by setting $A=1$. The other parameter values used in both simulations are $B=0.45$, $\nu=50$, $d=100$, $a=2$, $C=4$ and the number of space points is $M=2^{9}$}\label{fig:Numerics nonlocal solution}
\end{figure}

In the analysis performed in previous sections, we considered the model on an infinite domain. To mimic this in the simulations, we will consider a subdomain centred in a larger domain with the following initial conditions; outside the smaller subdomain the system's initial state will be set to the steady state, while on the subdomain a random perturbation will be added. The idea of this is to choose the outer domain large enough so that any conditions imposed on the boundary of this domain (which are set to be periodic in our simulations) do no affect the solution on the inner subdomain in the finite time that is considered in the simulation. The solution is then only considered on the subdomain on which a perturbation was introduced. To solve the Integro-PDE system \eqref{eq:Intro Klausmeier nonlocal}, it is first transformed into an ODE system by discretising its space domain and then solved by the built-in MATLAB ODE solver ode15s. A significant simplification  is made by computing the convolution term using the fast Fourier transform, as it reduces the number of operations required to find the convolution from $O(M^2)$ to $O(M\log (M))$ in each step (e.g. \cite{Cooley1969}), where $M$ is the number of points of the space domain. Figure \ref{fig:Numerics nonlocal solution} shows typical solutions obtained by this method; in Figure \ref{fig:Numerics nonlocal solution no patterns} the rainfall was chosen large enough for the solution to converge to the steady state, while for Figure \ref{fig:Numerics nonlocal solution whole interval} parameters that produce a patterned solution of the nonlocal Klausmeier model using the Laplacian kernel were used.

Using these simulations, we set up a scheme, based on the amplitude of the oscillation of the solution of the nonlocal Klausmeier model \eqref{eq:Intro Klausmeier nonlocal} relative to the steady state that approximates the critical rainfall parameter $A_{\max}$, that is the maximum rainfall parameter supporting pattern formation, for different kernel functions $\phi(x)$. Unlike in the simulation results shown in Figure \ref{fig:Numerics nonlocal solution}, we run the simulations over a shorter amount of time (up to $t=30$), as we are only interested in the onset of spatial patterns rather than in any of their properties. The kernel functions used in our simulations are those introduced in Section \ref{sec:Intro}, i.e. the Laplacian \eqref{eq:Intro Laplacian kernel}, the Gaussian \eqref{eq:Intro Gaussian kernel} and the power law kernel \eqref{Intro:power law distribution kernel}. Note that the standard deviations $\sigma(a)$ are given by $\sigma(a) = \sqrt{2}/a$ for the Laplacian kernel, $\sigma(a_g) = 1/(\sqrt{2}\,a_g)$ for the Gaussian kernel and $\sigma(a_p) = \sqrt{2}/(\sqrt{b^2-5b+6}\,a_p)$ for the power law kernel, provided $b>3$. If the shape parameter of the power law kernel is $b\le 3$, its standard deviation is infinite and a meaningful comparison to other kernel functions cannot be performed based on their standard deviations. In our simulations we consider both $b=3.1$ and $b=4$.  As in previous sections, we will consider the case in which $C=2/\sigma(a)^2$, motivated by the limiting behaviour of the nonlocal model, and the cases in which either $C$ or $a$ is assumed to be constant and the other parameter is varied. 

\begin{figure}
	\centering
	\subfloat[\label{fig:Numerics C constant Amax plots}]{\includegraphics[width=0.49\textwidth]{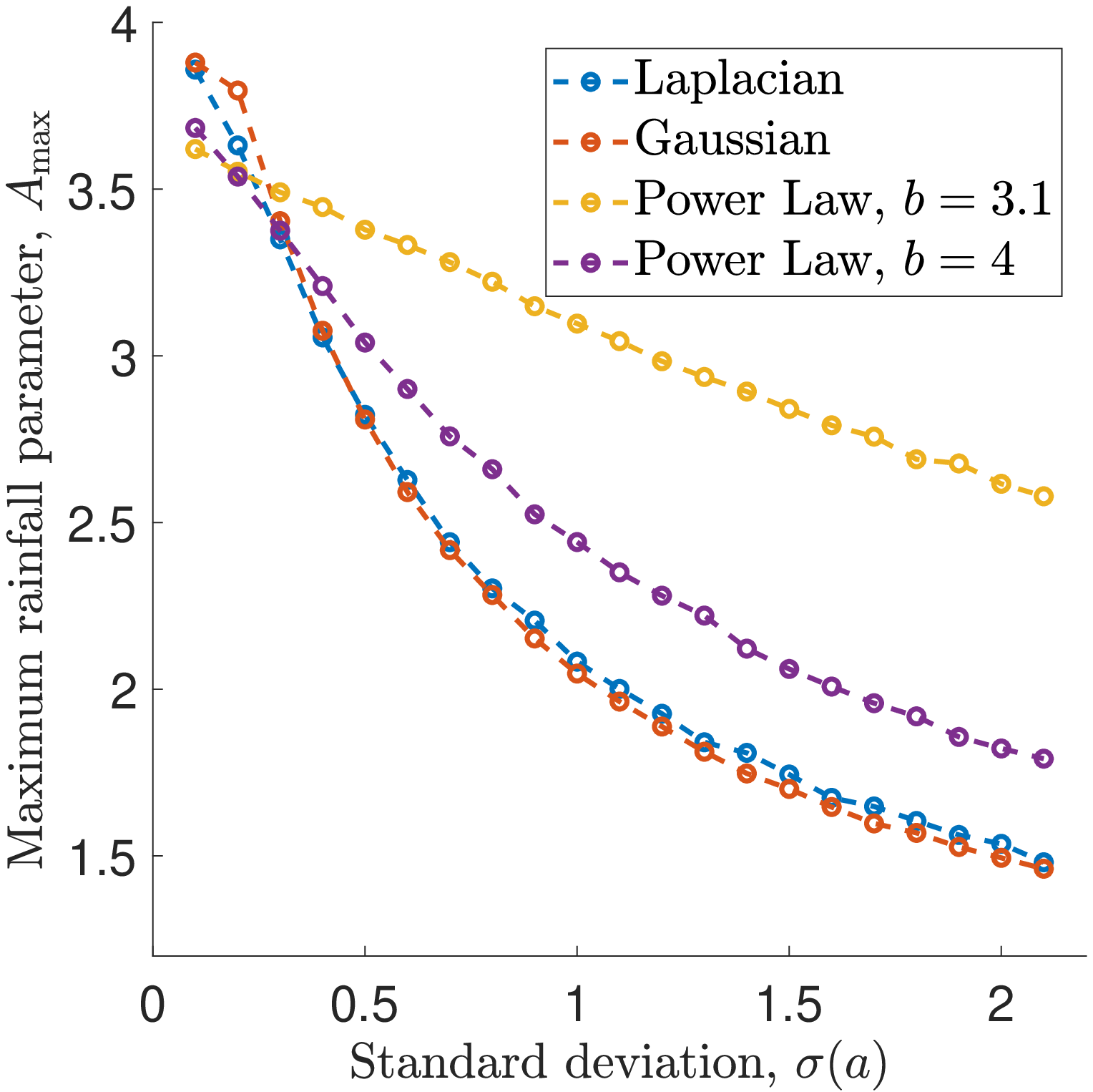}}
	\subfloat[\label{fig:Numerics C constant Amax comparison laplacian with leading order}]{\includegraphics[width=0.49\textwidth]{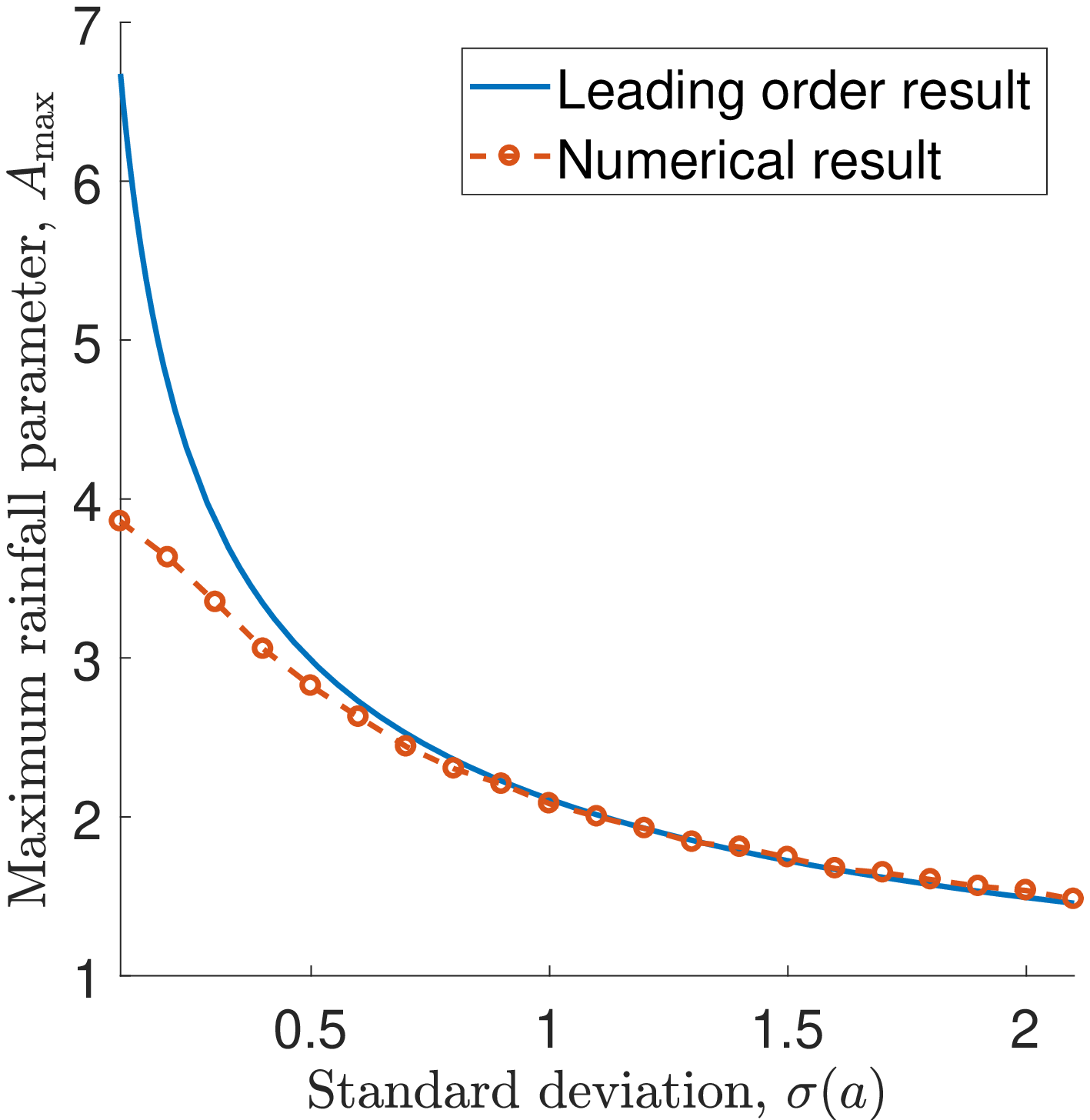}}
	\caption{Illustration of the results of our numerical scheme to approximate the maximum rainfall parameter $A_{\max}$ in the case of constant $C$. Part (a) shows the results of our simulations in the case of $C$ being constant. We have determined the maximum rainfall parameter giving patterns for the Laplacian kernel \eqref{eq:Intro Laplacian kernel}, the Gaussian kernel \eqref{eq:Intro Gaussian kernel} and the power law kernel \eqref{Intro:power law distribution kernel} for both $b=3.1$ and $b=4$, at $\sigma(a) = \{0.1,0.2, \dots 2.1\}$. The parameter values used in these simulations are $B=0.45$, $C=1$, $\nu = 50$, $d=1$. Part (b) compares the simulation results obtained for the Laplacian kernel to the corresponding condition \eqref{eq:Hopf condition on A Nonlocal d not 0} valid to leading order in $\nu$}
\end{figure}

Figure \ref{fig:Numerics C constant Amax plots} shows the results of our simulations in the case of $C$ being constant. The trend that a narrower dispersal kernel requires a higher level of rainfall to form homogeneous vegetation, which was predicted by the leading order form \eqref{eq:Hopf condition on A Nonlocal d not 0} of $A_{\max}$ for the Laplacian kernel, carries over to the other kernels used in the simulations. Further one can observe that the power law distributions which have algebraic decay give a larger value of $A_{\max}$ than those with exponential decay if the standard deviation is sufficiently large ($\sigma(a) \gtrapprox 0.3$), while for narrower kernels the opposite is true. While the results of our simulations for the Laplacian kernel and the corresponding leading order form of $A_{\max}$ fit well for sufficiently large values of the standard deviation $\sigma(a)$, the fit is poorer for narrower kernel functions (see Figure \ref{fig:Numerics C constant Amax comparison laplacian with leading order} for a comparison). The reason for this is the relatively small choice of $\nu=50$, which was taken to improve the speed of the simulations. Solutions for larger $\nu$ indicate that the relative difference between $A_{\max}$ in our simulations and in the analytical approximation decreases (slowly) with increasing $\nu$.

\begin{figure}
	\centering
	\includegraphics[width=0.50\textwidth]{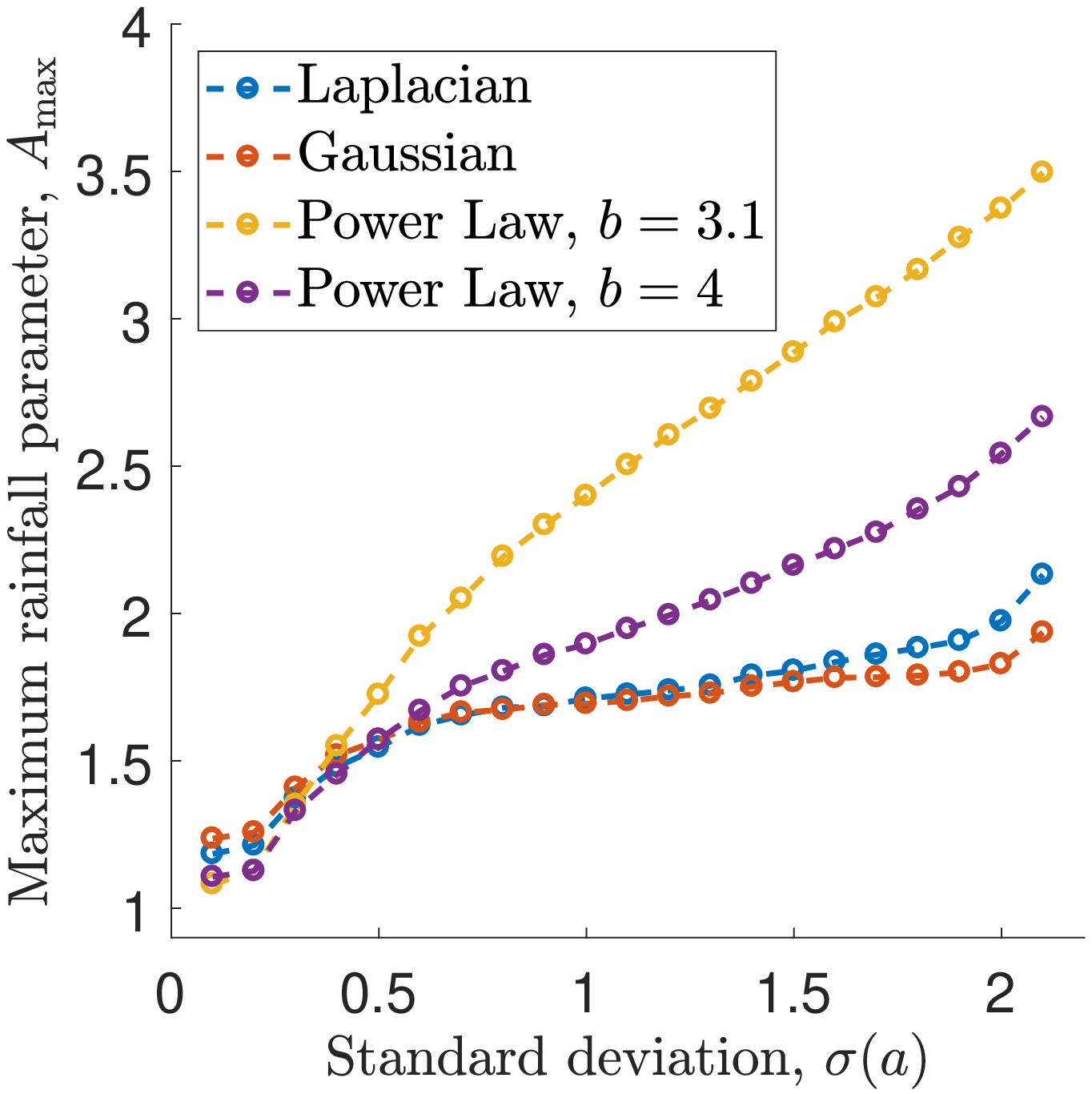}
	\caption{Illustration of the results of our numerical scheme to approximate the maximum rainfall parameter $A_{\max}$ in the case of $C = 2/\sigma(a)^2$. We have considered the Laplacian kernel \eqref{eq:Intro Laplacian kernel}, the Gaussian kernel \eqref{eq:Intro Gaussian kernel} and the power law kernel \eqref{Intro:power law distribution kernel} for both $b=3.1$ and $b=4$ and determined the value of the maximum rainfall parameter giving patterns $A_{\max}$ at $\sigma(a) = \{0.1,0.2, \dots 2.1\}$ for each kernel function in the case of $C = 2/\sigma(a)^2$. The parameters used in this simulation are $B=0.45$, $\nu = 50$, $d=1$}\label{fig:Numerics C a squared Amax}
\end{figure}

We repeat the same scheme in the setting of $C=2/\sigma(a)^2$ for the same kernel functions. The results of this are shown in Figure \ref{fig:Numerics C a squared Amax}. Considering the type of decay of the kernel functions, the results of it are similar to the simulations of the case of $C$ being constant. One can observe that the distributions with algebraic decay yield a larger value of $A_{\max}$ than the distributions with exponential decay if the kernel is sufficiently wide, while for narrow kernels the opposite is true. Further, considering one specific kernel on its own, a narrower dispersal kernel now gives a lower value of the maximum rainfall parameter supporting pattern formation. This is in contrast to the case in which $C$ was kept constant but in accord with the leading order form \eqref{eq:Hopf condition on A Nonlocal d not 0} of $A_{\max}$. As before, it can also be observed from the simulations of the model using the Laplacian kernel that for the choice of $\nu=50$, the numerical simulations are a good approximation of the leading order result only for sufficiently wide kernels.

\begin{figure}
	\centering
	\subfloat[\label{fig:Numerics Amax a fixed large sd}]{\includegraphics[width=0.49\textwidth]{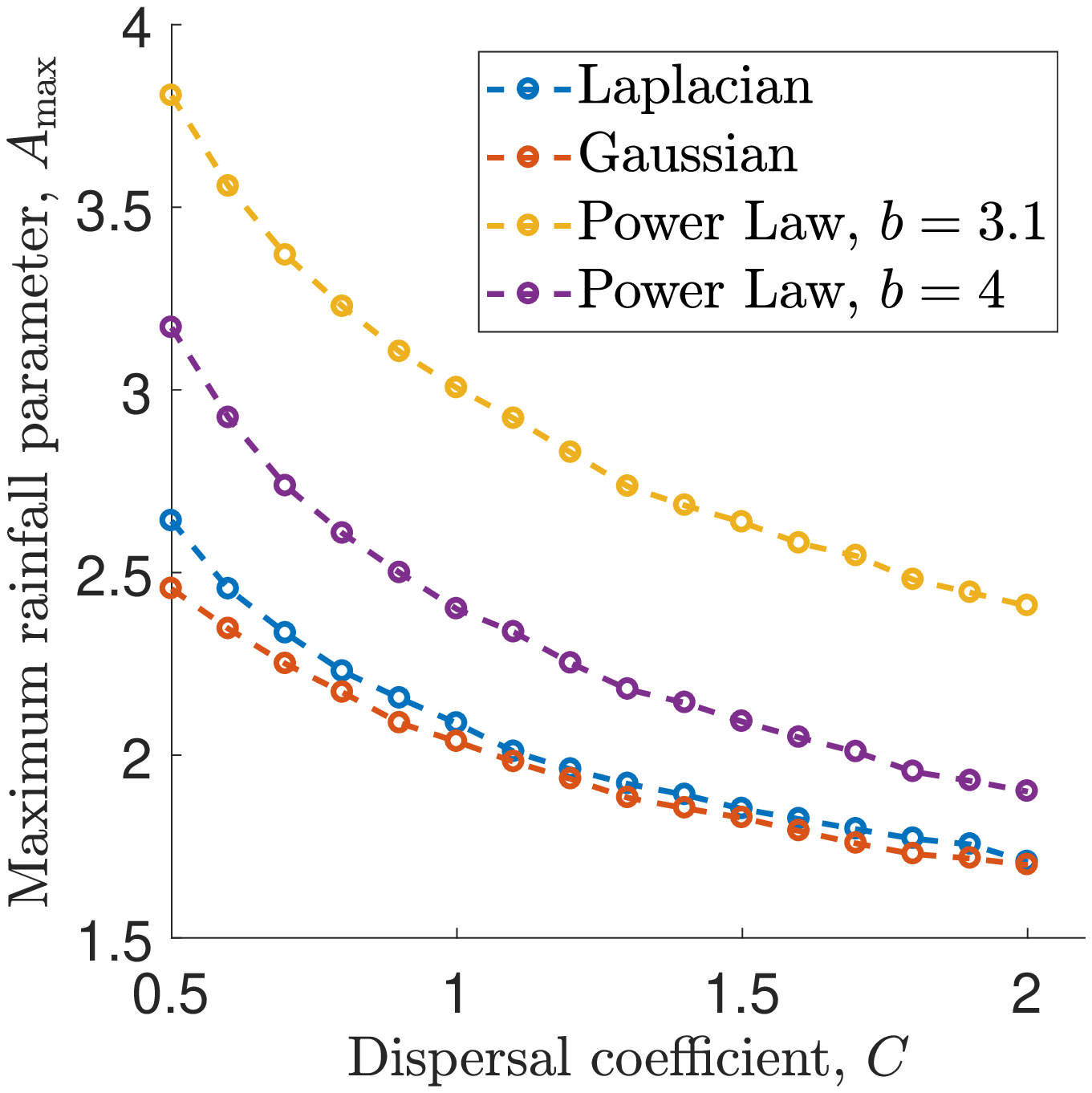}}
	\subfloat[\label{fig:Numerics Amax a fixed small sd}]{\includegraphics[width=0.49\textwidth]{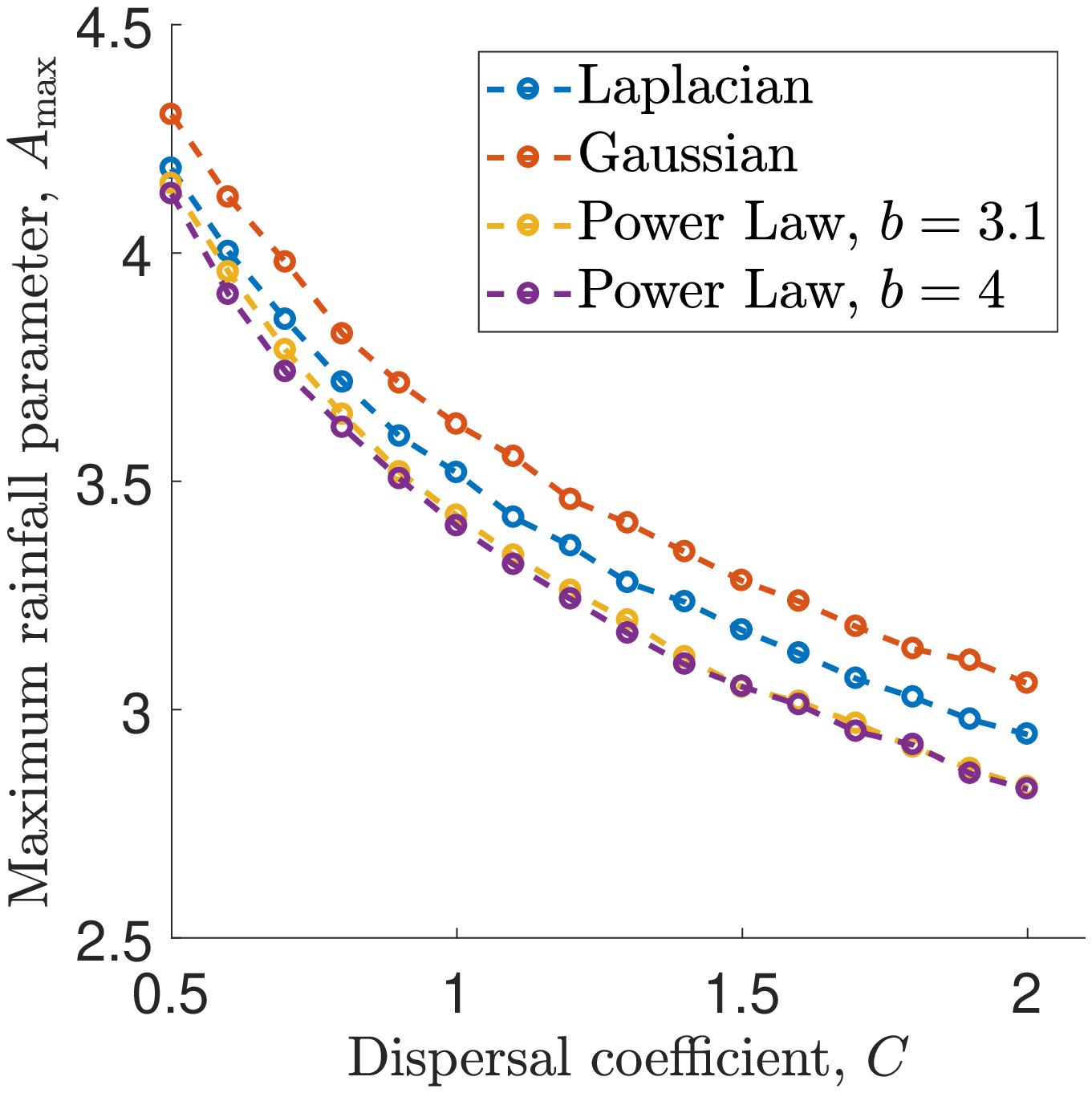}}
	\caption{Illustration of the results of our numerical scheme to approximate the maximum rainfall parameter $A_{\max}$ in the case of constant $a$. We have considered the Laplacian kernel \eqref{eq:Intro Laplacian kernel}, the Gaussian kernel \eqref{eq:Intro Gaussian kernel} and the power law kernel \eqref{Intro:power law distribution kernel} for both $b=3.1$ and $b=4$ and determined the value of the maximum rainfall parameter giving patterns $A_{\max}$ at $C = \{0.5,0.6, \dots 2\}$ for each kernel function, with $a$ being fixed. In (a), the standard deviation was chosen as $\sigma(a)=1$, in (b) as $\sigma(a) = 0.2$. The other parameters used in this simulation are $B=0.45$, $\nu = 50$, $d=1$}\label{fig:Numerics Amax a fixed}
\end{figure}

Finally, we apply the same scheme to the case of fixed dispersal parameter $a$ and varying dispersal coefficient $C$ (Figure \ref{fig:Numerics Amax a fixed}). As in the previous cases, the trends of the simulations of other kernel functions are again in alignment with the leading order result \eqref{eq:Hopf condition on A Nonlocal d not 0} for the Laplacian kernel. An increase in dispersal rate $C$ causes a decrease in $A_{\max}$ for each of the dispersal kernels considered in our simulations. Further, the comparison of kernels with algebraic and exponential decay depends on the choice of standard deviation $\sigma(a)$, as indicated by the previous simulations in which the standard deviation was varied. Figure \ref{fig:Numerics Amax a fixed small sd} shows that for a small standard deviation ($\sigma(a)=0.2$ in this case), the kernels with exponential decay predict that a higher level of rainfall is required to form a uniform vegetation cover than those with algebraic decay. If the standard deviation is sufficiently large, the opposite trend is observed. This is visualised in Figure \ref{fig:Numerics Amax a fixed large sd}, where the standard deviation was set to $\sigma(a)=1$.


\section{Discussion}\label{sec:Comparison}

The main results of this paper are given by \eqref{eq:Hopf condition on c Nonlocal d not 0} and \eqref{eq:TWS condition on c that omega real, d not 0}, which give an upper bound for the parameter region in the $A$-$c$ plane supporting pattern formation, valid to leading order in $\nu$, for the nonlocal Klausmeier model \eqref{eq:Intro Klausmeier nonlocal} with the Laplacian kernel \eqref{eq:Intro Laplacian kernel}. In particular this gives the upper bound $A_{\max}$, defined in \eqref{eq:Hopf condition on A Nonlocal d not 0}, on the rainfall parameter, again valid to leading order in $\nu$. In other words, $A_{\max}$ represents the lowest level of rainfall that allows plants to form a homogeneous vegetation cover, while lower amounts of water only support banded vegetation. These results hold under the assumptions that the migration speed $c$ is $O_s(1)$ and that the parameter region supporting pattern formation is bounded above by the loci of Hopf bifurcations, which was shown by \cite{Sherratt2007} for the local Klausmeier model \eqref{eq:Intro Klausmeier local}. While the simple nature of the Klausmeier model makes it impossible to deduce any quantitative conclusions from these results, they do give a good insight into the parametric trends of the model. These trends fundamentally depend on the assumption made on the factor $C$ scaling the convolution term in the nonlocal model. 

In this paper we considered three different cases of the coefficient $C$ in the nonlocal Klausmeier model; that of choosing it to be constant, the one of varying $C$ for fixed dispersal parameter $a$ and that of setting $C=2/\sigma(a)^2$. In the case of $C$ being fixed, a change in the dispersal parameter $a$ only affects the width of the dispersal kernel, but leaves the term scaling the nonlocal plant dispersal term unchanged. It can be immediately concluded from \eqref{eq:Hopf condition on A Nonlocal d not 0} that the threshold $A_{\max}$ increases as the kernel width decreases. This increase in the size of the parameter region supporting pattern formation is also visualised in Figure \ref{fig:TWS Ac plane nonlocal C fixed}. This means that the wider plants disperse their seeds, the less water they require to form a homogeneous vegetation cover. In particular, our results show that if plant dispersal is wide enough, the location of the Hopf bifurcation bounding the pattern forming parameter region completely lies in the region that only supports the trivial steady state describing complete desertification. In this case, the assumptions taken in this paper predict that no striped vegetation can occur. Plants either form a homogeneous vegetation cover or disappear completely. 

The expression given by \eqref{eq:Hopf condition on A Nonlocal d not 0} is only valid for the Laplacian kernel \eqref{eq:Intro Laplacian kernel} and to leading order in $\nu$. The numerical simulations in Section \ref{sec:Numerics} allow us to compare this condition to those for other kernel functions that have been suggested by studies on plant dispersal (see \cite{Bullock2017} for an overview). Our results suggest that the maximum rainfall level giving patterned vegetation depends on the width and therefore also on the type of decay of the dispersal kernel. It can be seen from Figures \ref{fig:Numerics C constant Amax plots} and \ref{fig:Numerics C a squared Amax} that those probability distributions that decay algebraically predict a larger pattern-giving parameter region for some fixed standard deviation than those decaying exponentially under all the different assumptions taken on $C$ in this paper, if the dispersal kernel is sufficiently wide. If the kernel is narrow, the opposite behaviour is observed. Further, the simulations show that $A_{\max}$ for each individual kernel is decreasing as the width of the kernel is increased if one assumes that $C$ is constant. This is in accord with the behaviour of the leading order form \eqref{eq:Hopf condition on A Nonlocal d not 0} of the Laplacian kernel. Combining these observations, we can conclude that the narrower a plant's seed dispersal is, the more water is required to avoid the formation of patterns. Nevertheless, field data shows that plants in semi-arid ecosystems tend to establish narrow dispersal kernels \cite{Ellner1981, RheedevanOudtshoorn2013}. This is, however, only a side effect of other adaptations such as seed containers protecting seeds from flooding and predation \cite{Ellner1981}. Simulations show that short range dispersal yields a higher mean biomass in those ecosystems than a long distance spread of seeds \cite{Pueyo2008}. Combining this with the results of this paper shows that the shortening of dispersal ranges of plants in semi-arid environments increases their tendency to self-organise into patterns.

If one assumes that the width of the dispersal kernel is fixed and plant's dispersal rate is changed, \eqref{eq:Hopf condition on A Nonlocal d not 0} shows that, under the assumption that the dispersal of seeds fits the Laplacian kernel, the more the species invests in its dispersal rate, the less water it requires to form a homogeneous vegetation cover. For the other dispersal kernels we have considered, the same behaviour is shown in our simulations. Those simulations also show the same trend regarding the type of decay of the dispersal kernels as the simulations in the case of fixed $C$ and varying range of dispersal. For wider dispersal kernels, those plants whose kernel functions decay algebraically have a higher tendency to form patterns than those plants dispersing their seeds according to an exponentially decaying kernel. For sufficiently narrow kernels, the opposite observation can be made.

The final choice of $C$ assumes that it is correlated with the standard deviation of the dispersal kernel as $C=2/\sigma(a)^2$. This choice is of particular significance because it leads to the local Klausmeier model being a limiting case of the nonlocal model using either the Laplacian or the Gaussian kernel. This allows us to compare our results to the corresponding results obtained for the local model by \cite{Sherratt2013IV}. This choice is motivated purely mathematically and we are not aware of any evidence that the dispersal coefficient $C$ is correlated with the seed distribution range in such a way. However, experiments have shown that plants' rate of dispersal increases in semi-arid environments \cite{Aronson1993}, e.g. by the production of more but smaller seeds \cite{Volis2007} as well as that plants develop short range dispersal of seeds \cite{Ellner1981, RheedevanOudtshoorn2013}. The analysis of the previous two cases has shown that an increase in the dispersal coefficient reduces the critical level of rainfall required to form a homogeneous vegetation cover, while the establishment of a narrow dispersal kernel increases this threshold. Therefore, this could be seen as an evolutionary trade-off.

\begin{figure}
	\centering
	\includegraphics[width=0.6\textwidth]{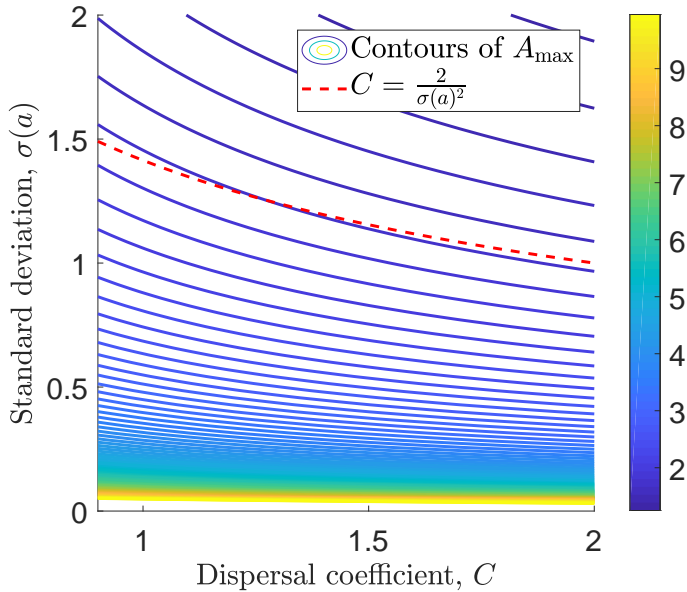}
	\caption{Contour plot of $A_{\max}$. This plot shows the contours of \eqref{eq:Hopf condition on A Nonlocal d not 0} as solid lines with the colours indicating the level of $A_{\max}$. The red dotted line is the suggested trade-off $C=2/\sigma(a)^2$, which was mathematically motivated by the limiting behaviour of the convolution integral}\label{fig:Discussion Motivation C=a^2}
\end{figure}

The leading order results on quantities such as $A_{\max}$ or the wavelength, obtained in Sections \ref{sec:Linear Stability} and \ref{sec:TWS}, resemble the limiting behaviour of this case. Apart from the limiting case, the results for the nonlocal model using the Laplacian kernel behave monotonically as the width of the dispersal kernel is changed. In particular, the results on the loci of the Hopf bifurcation and the maximum rainfall parameter giving patterns allow us to make the crucial observation that the nonlocal model predicts a larger range of parameters supporting pattern formation. Our results further show that the size of the parameter region giving patterns is larger for a wider dispersal kernel, which makes the dispersal term less influential, i.e. it decreases the plant's dispersal rate, due to the assumption $C=a^2$. This is most strikingly illustrated by Figure \ref{fig:TWS Ac plane nonlocal comparison}, which shows the increase of this region as the scale parameter $a$ of the kernel decreases. Under this assumption on the dispersal rate and the kernel width, our simulation results show that establishing short range dispersal increases plants' ability to form a homogeneous vegetation cover. This is further illustrated by Figure \ref{fig:Discussion Motivation C=a^2}, which shows shows the contours of $A_{\max}$ and the suggested evolutionary trade-off $C=2/\sigma(a)^2$. The latter crosses the contours as the standard deviation is varied and thereby shows that an increase in kernel width yields an increase in the maximum rainfall parameter supporting pattern formation.

In this paper we have also investigated the distance between the striped vegetation patches to leading order in $\nu$. It is of immense importance to have an understanding of the wavelength of the patterns as it might give an indication of whether the ecosystem is close to complete desertification. The results of this study show that the wavelength monotonically increases as the amount of rainfall decreases, before reaching a critical threshold, where patterns disappear and complete desertification takes over. While it is important to emphasise again that the simplifications assumed in deducing the Klausmeier model do not allow us to gain any quantitative information, we have shown how the wavelength is affected by changes in the width of the dispersal kernel or in the plant's dispersal rate. Interestingly, in the case of $C=2/\sigma(a)^2$, the wavelength predicted by the nonlocal model using the Laplacian kernel does not differ much from the wavelength predicted by the local model, even for wide dispersal kernels (see the $y$-axis in Figure \ref{fig:Wavelength comparison}). This suggests that one could make predictions on the possibility of desertification without having any information on the range of plant dispersal under the assumption that the dispersal coefficient $C$ is correlated with the standard deviation of the dispersal kernel in such a way.

The pattern solutions of the Klausmeier model fundamentally depend on how the migration speed $c$ scales with the parameter $\nu$, describing the rate of the water flow downhill. In this paper we have only considered the case $c=O_s(1)$ and the patterns forming in the vicinity of the Turing-Hopf bifurcation. For the local Klausmeier model results have been obtained for a wide range of migration speeds \cite{Sherratt2010, Sherratt2011, Sherratt2013III, Sherratt2013IV, Sherratt2013V}. One natural extension of this work would be to do a similar comprehensive study of the whole parameter range for the nonlocal model. This would give insights into the existence and form of patterns away from the bifurcation point.

Another natural area for future work would be to consider other more realistic models for vegetation patterns. A number of such models and their underlying mechanisms and scale dependent feedbacks are reviewed by \cite{Meron2012}. Some of these models already include nonlocal dispersal via convolution integrals \cite{Baudena2013, Pueyo2008, Pueyo2010}, and in others \cite{HilleRisLambers2001, Rietkerk2002} such a term could be added in place of plant diffusion. While these as well as the model considered in this paper assume an isotropic dispersal of plants, this simplification can be removed by including either advection of plants \cite{Thompson2009, Saco2007} or an asymmetric dispersal kernel \cite{Thompson2009}. Similar to the relation between diffusion and the convolution with a symmetric kernel, both the advection and the diffusion terms arise from the convolution term with an asymmetric kernel. In this case the coefficient of the first order derivative in \eqref{eq: intro: convolution expansion} is non-zero. Finally, some models use a nonlocal term for the water uptake and thus also for plant growth, reflecting the extensive root networks of plants in semi-arid regions \cite{Gilad2004, Gilad2007}. Investigation of these models using an approach similar to that in the current paper would be of interest but would be particularly challenging because of the added complexity.

\section*{Acknowledgements}
Lukas Eigentler was supported by The Maxwell Institute Graduate School in Analysis and its Applications, a Centre for Doctoral Training funded by the UK Engineering and Physical Sciences Research Council (grant EP/L016508/01), the Scottish Funding Council, Heriot-Watt University and the University of Edinburgh.

\clearpage
\bibliography{bibliography.bib}

\end{document}